\documentclass{article}

\usepackage{arxiv}

\usepackage{graphicx}
\usepackage{wrapfig}
\usepackage{float}
\usepackage{enumitem}
\usepackage[dvipsnames,table]{xcolor}
\usepackage{amssymb}
\usepackage{amsmath}
\usepackage{amsfonts}       
\usepackage{bm}
\usepackage{bbm}
\usepackage{multirow}
\usepackage{booktabs}       
\usepackage{microtype}      
\usepackage{nicefrac}       
\usepackage[utf8]{inputenc} 
\usepackage[T1]{fontenc}    

\usepackage[linesnumbered,ruled,vlined]{algorithm2e}
\usepackage{algorithmic}
\usepackage{varwidth}

\usepackage[colorlinks=true,
linkcolor=RoyalBlue,
urlcolor=RoyalBlue,
citecolor=RoyalBlue,
anchorcolor=RoyalBlue,
backref=page]{hyperref}       
\usepackage{url}            
\usepackage{orcidlink}
\newcommand{\ORCID}[1]{\href{https://orcid.org/#1}{\orcidlink{#1}\,https://orcid.org/#1}}

\makeatletter
\renewcommand{\numberline}[1]{%
  \hb@xt@2.5em{#1\hfil}%
}
\makeatother

\title{Scalable Multitemperature Free Energy Sampling of Classical Ising Spin States}

\author{
Ping Tuo \\
Bakar Institute of Digital Materials for the Planet,
University of California, Berkeley,
CA 94720, United States \\
Institute of Science and Technology Austria,
3400 Klosterneuburg, Austria \\
\ORCID{0000-0002-6477-5900} \\
Email: \href{mailto:tuoping@berkeley.edu}{tuoping@berkeley.edu}
\And
Zezhu Zeng \\
Institute of Science and Technology Austria,
3400 Klosterneuburg, Austria \\
\ORCID{0000-0001-5126-4928}
\And
Jiale Chen \\
Institute of Science and Technology Austria,
3400 Klosterneuburg, Austria \\
\ORCID{0000-0001-5337-5875}
\AND
Bingqing Cheng \\
Department of Chemistry,
University of California, Berkeley,
CA 94720, United States \\
Institute of Science and Technology Austria,
3400 Klosterneuburg, Austria \\
\ORCID{0000-0002-3584-9632}
}

\begin{document}
\maketitle

\begin{figure}[h]
\centering
\includegraphics[width=0.682\linewidth]{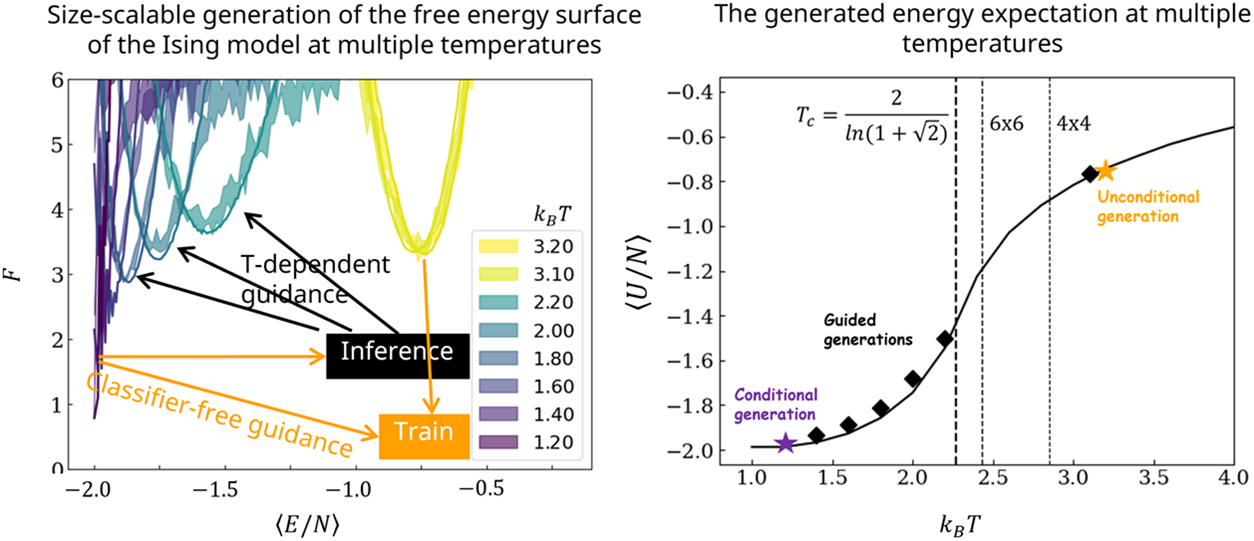}
\end{figure}

\begin{abstract}
Generative models have advanced significantly in sampling material systems with continuous variables, such as atomistic structures. However, their application to discrete variables, like atom types or spin states, remains underexplored. 
In this work, we introduce a discrete flow matching model, tailored for systems with discrete phase-space coordinates (e.g., the Ising model or a multicomponent system on a lattice). This approach enables a single model to sample free energy surfaces over a wide temperature range with minimal training overhead, and the model generation is scalable to larger lattice sizes than those in the training set.
We demonstrate our approach on the 2D Ising model, showing efficient and reliable free energy sampling. 
These results highlight the potential of flow matching for low-cost, scalable free energy sampling in discrete systems and suggest promising extensions to alchemical degrees of freedom in crystalline materials. 
The codebase developed for this work is openly available at \url{https://github.com/tuoping/alchemicalFES}.
\end{abstract}

\clearpage

\tableofcontents

\clearpage

\section{Introduction}
Estimating the free energy surface (FES) of the alchemical space of crystalline solids with different elements,
which is isomorphic to an Ising spin system or a lattice model,
has traditionally relied on stochastic sampling methods, such as Markov chain Monte Carlo (MCMC) simulations~\cite{frenkel2023understanding}.
MCMC methods sample by constructing a Markov chain whose equilibrium distribution matches the target distribution, with common approaches like Metropolis-Hastings~\cite{hastings1970monte}, simulated annealing~\cite{kirkpatrick1983optimization}, and replica exchange~\cite{marinari1992simulated,hukushima1996exchange} helping to overcome challenges like metastability and slow convergence. These methods sample from the Boltzmann distribution in the long run, but many simulation steps are needed to produce an uncorrelated sample. One reason is that complex systems often have metastable states and the transitions between them are rare events. For instance, for the Ising model on a 24 $\times$ 24 square lattice at temperature $T=0.88\ T_c$ (below the critical temperature $T_c$), more than $10^9$ MCMC steps are required to flip the overall magnetization direction; at a lower temperature $T=0.79\ T_c$, the flipping failed to happen after $10^{12}$ MCMC steps.

Recently, deep generative models, such as normalizing flows, diffusion, and flow matching, have emerged as promising methods for estimating free energy surfaces (FES). By mapping the complex configurational space with a Boltzmann distribution to a latent space, in which the low-energy configurations of different states lie close to each other, these models enable more efficient sampling~\cite{noe2019boltzmann,klein2024equivariant,causer2024discrete}. For instance, Wang et al.~\cite{wang2022data} used diffusion models to reduce the number of replicas required in replica exchange schemes. Olehnovics et al.~\cite{olehnovics2024assessing} employed normalizing flows for more efficient reweighting in targeted free energy perturbation. Olsson et al.~\cite{moqvist2024thermodynamic} used a flow matching model to simulate thermal interpolation. Herron et al.~\cite{herron2024inferring} trained a diffusion model to generate temperature-dependent attributes. While Dibak et al.~\cite{dibak2022temperature} trained a temperature steerable normalizing flow to generate temperature-dependent attributes. Together, these examples underscore the versatility and potential impact of deep generative modeling for advancing the study of complex free energy landscapes.

Although substantial progress has been made in applying generative models to continuous spaces, their use in discrete systems, such as spin lattices, remains comparatively limited. In particular, for the Ising model where the spins take values in $\{-1, +1\}$, the continuous-space formulations do not apply.
A diverse but still emerging set of approaches has been developed, including masked modeling~\cite{ghazvininejad2019mask,chang2022maskgit}, autoregressive models~\cite{wu2019solving,sharir2020deep,singha2025multilevel}, discrete diffusion models~\cite{austin2021structured,hoogeboom2021argmax,campbell2024trans,causer2024discrete,lou2023discrete,avdeyev2023dirichlet,han2022ssd}, and discrete flow matching~\cite{stark2024dirichlet,gat2024discrete,campbell2024generative,zhao2024probabilistic,davis2024fisher,miller2024flowmm,nicoli2020asymptotically,nicoli2021estimation,bulgarelli2024flow}.
Masked modeling techniques use multiple iterations of masking and filling to gradually reconstruct the entire input~\cite{ghazvininejad2019mask,chang2022maskgit}. Autoregressive models generate discrete sequences element by element, capturing strong dependencies~\cite{wu2019solving,sharir2020deep,singha2025multilevel}. They can be effective in many settings, though scaling to high-dimensional data requires careful consideration. Discrete diffusion models extend denoising diffusion from continuous to categorical data~\cite{austin2021structured,hoogeboom2021argmax,campbell2024trans,causer2024discrete,lou2023discrete,avdeyev2023dirichlet,han2022ssd}, leveraging iterative sampling steps to reconstruct complex distributions. In contrast, discrete flow matching learns a deterministic transformation from a simple distribution to the target distribution without iterative noise addition and removal, offering a faster inference process compared to iterative approaches~\cite{stark2024dirichlet,gat2024discrete,campbell2024generative,zhao2024probabilistic,davis2024fisher,miller2024flowmm}.

In this work, we developed alchemicalFES, a flow matching model in the alchemical space. The model exhibits two key advancements in transferability: (1) it generates the FES across multiple temperatures using one trained model; (2) after being trained on the MCMC data of a small lattice, the model is scalable to lattices of arbitrary sizes. In Section~\ref{sec:alchemicalfes}, we detail the specific FM formulations used in this work.
After describing the algorithm, in Section~\ref{sect:fes-ising-uncond}, we apply the model to generate the FES of a 2D square-lattice Ising model with the Hamiltonian $H = -\frac{1}{2}\sum_{i=1}^N \sum_{j=1}^N J_{ij} s_i s_j$ where $N$ is the number of lattice sites, $s_i \in \{-1, +1\}$, and $J_{ij} = 1$ if sites $i$ and $j$ are nearest neighbors and $0$ otherwise. Finally, in Section~\ref{sect:multi-T-FM}, we show how to achieve multitemperature generation by using the classifier-free guidance technique~\cite{ho2022classifier}.

\section{alchemicalFES Architecture}\label{sec:alchemicalfes}
We begin by reviewing foundational work on flow matching in Section~\ref{sec:fm}, where a neural network is employed to simulate an ordinary differential equation. In Section~\ref{sec:FM-simplex}, we extend this framework to flow matching based on Dirichlet distributions, a formulation naturally suited for discrete variables. The neural network architecture developed in this work is presented in Section~\ref{sec:cnn}. Finally, Section~\ref{sec:FM-inference} details how inference is performed through integration.

\subsection{Definition of Flow Matching}\label{sec:fm}

In flow matching (FM), let the time-indexed random vector $\boldsymbol{x} \left( t \right) \in \mathbb{R}^d$ for $t \in \left[ 0, 1 \right]$ have density $p_t \left( \boldsymbol{x} \right)$ over $\boldsymbol{x} \in \mathbb{R}^d$.
Let $q \left( \boldsymbol{x} \right)$ denote the source/noise density and $p_{\mathrm{data}} \left( \boldsymbol{x} \right)$ the target/data density; we impose the boundary conditions $p_0 = q$ and $p_1 = p_{\mathrm{data}}$.
We introduce a time-evolving conditional density $p_t \left( \boldsymbol{x} \mid \boldsymbol{x} \left( 1 \right) \right)$ of $\boldsymbol{x} \left( t \right)$ conditioned on $\boldsymbol{x} \left( 1 \right)$ with
$p_0 \left( \boldsymbol{x} \mid \boldsymbol{x} \left( 1 \right) \right) = q \left( \boldsymbol{x} \right)$
and
$p_1 \left( \boldsymbol{x} \mid \boldsymbol{x} \left( 1 \right) \right) = \delta \left( \boldsymbol{x} - \boldsymbol{x} \left( 1 \right) \right)$,
where $\delta$ is the Dirac delta function.
The corresponding marginal density is the mixture
\begin{equation}
p_t \left( \boldsymbol{x} \right)
= \int_{\mathbb{R}^d} p_t \left( \boldsymbol{x} \mid \boldsymbol{x} \left( 1 \right) \right) p_{\mathrm{data}} \left( \boldsymbol{x} \left( 1 \right) \right) \mathrm{d}\boldsymbol{x} \left( 1 \right) .
  \label{eq:marg-pp}
\end{equation}
For each $\boldsymbol{x} \left( 1 \right)$, let $\boldsymbol{u}_t \left( \boldsymbol{x} \mid \boldsymbol{x} \left( 1 \right) \right) \in \mathbb{R}^d$ be a conditioned velocity field transporting $p_t \left( \boldsymbol{x} \mid \boldsymbol{x} \left( 1 \right) \right)$ via the continuity equation
\begin{equation}
\frac{\partial}{\partial t} p_t \left( \boldsymbol{x} \mid \boldsymbol{x} \left( 1 \right) \right)
+ \nabla_{\boldsymbol{x}} \cdot \left( p_t \left( \boldsymbol{x} \mid \boldsymbol{x} \left( 1 \right) \right) \boldsymbol{u}_t \left( \boldsymbol{x}\mid \boldsymbol{x} \left( 1 \right) \right) \right) = 0 .
\label{eq:transport}
\end{equation}
The induced marginal velocity field is the posterior average
\begin{equation}
\boldsymbol{v}_t \left( \boldsymbol{x} \right)
= \int_{\mathbb{R}^d} \boldsymbol{u}_t \left( \boldsymbol{x} \mid \boldsymbol{x} \left( 1 \right) \right)\,
\frac{p_t \left( \boldsymbol{x} \mid \boldsymbol{x} \left( 1 \right) \right)  p_{\mathrm{data}} \left( \boldsymbol{x} \left( 1 \right) \right)}{p_t \left( \boldsymbol{x} \right)} 
\mathrm{d}\boldsymbol{x} \left( 1 \right) .
\label{eq:marg-vf}
\end{equation}
In practice, we train a neural network to approximate $\boldsymbol{v}_t \left( \boldsymbol{x} \right)$.
Sampling is then obtained by integrating the deterministic flow
\begin{equation}
\frac{\mathrm{d}}{\mathrm{d} t} \boldsymbol{x}\left( t \right) = \boldsymbol{v}_t \left( \boldsymbol{x} \left( t \right) \right) .
\label{eq:ode-sampling}
\end{equation}
Thus, we can generate data $\boldsymbol{x} \left( 1 \right) \sim p_\mathrm{data} \left( \bm{x} \right)$ from noisy samples $\boldsymbol{x} \left( 0 \right) \sim q \left( \bm{x} \right)$.

\subsection{Flow Matching on the Simplex}\label{sec:FM-simplex}

For an Ising model, each spin $s_i$ at the $i$-th lattice site can take one of two states: $\left\{ -1, +1 \right\}$. We represent these two states with a categorical distribution, using probabilities $\boldsymbol{x}_i= \left( x_{i(0)}, x_{i(1)} \right)$ that satisfy $x_{i(0)} + x_{i(1)} = 1$, $x_{i(0)}, x_{i(1)} \ge 0$. Then, each spin state can take two values $\boldsymbol{x}_i(t=1)=(1,0)$ or $(0,1)$. To describe the distribution of $\boldsymbol{x}_i$, we use the Dirichlet distribution, which has the probability density function~\cite{kotz2019continuous}
\begin{equation}
p \left( \boldsymbol{x}_i; \boldsymbol{\alpha} \right) 
= \text{Dir} \left( \boldsymbol{x}_i; \left( \alpha_0, \alpha_1 \right) \right)
= \frac{\Gamma \left( \alpha_0 \right) \Gamma \left( \alpha_1 \right)}{\Gamma \left( \alpha_0 + \alpha_1 \right)}
x_{i(0)}^{\alpha_0 - 1} x_{i(1)}^{\alpha_1 - 1}
\label{eq:binary-dir}
\end{equation}
with parameters $\boldsymbol{\alpha} = \left( \alpha_0, \alpha_1 \right)$, $\alpha_0, \alpha_1 > 0$ , and the gamma function $\Gamma \left(\cdot\right)$~\cite{davis1959leonhard}. 

We then define the noisy prior $q$ to be the uniform distribution, or a \emph{Dirichlet distribution} with parameter $\boldsymbol{\alpha}=\left(1,1\right)$:
\begin{equation}
    q \left( \boldsymbol{x} \right) = \text{Dir}\left( \boldsymbol{x}; \boldsymbol{\alpha} = \left( 1, 1 \right) \right) = 1.
    \label{eq:dirbasis}
\end{equation}
See Supporting Figure~\ref{fig:conv-dirichlet}(a) for an illustration of the Dirichlet distribution with different $\boldsymbol{\alpha}$ parameters.
Then, we define a conditional probability path by increasing one of the entries of $\boldsymbol{\alpha}$ with time $t \in \left[ 0, 1 \right]$:
\begin{align}
\nonumber p_t \left( \boldsymbol{x} \mid \boldsymbol{x}(1) \right) = & 
\mathrm{Dir} \left( \boldsymbol{x}; \boldsymbol{\alpha} = \left( 1, 1 \right) + t\alpha_\mathrm{max} \cdot \boldsymbol{x}(1) \right) \\ 
= & \begin{cases}
    \mathrm{Dir} \left( \boldsymbol{x}; \boldsymbol{\alpha} = \left( 1 + t \alpha_\mathrm{max}, 1 \right) \right), & \text{if } \boldsymbol{x}(1) = \left( 1, 0 \right) \\
    \mathrm{Dir} \left( \boldsymbol{x}; \boldsymbol{\alpha} = \left( 1, 1 + t \alpha_\mathrm{max} \right) \right), & \text{if } \boldsymbol{x}(1) = \left( 0, 1 \right)
\end{cases}
\label{eq:probpath-dir}
\end{align}
where $\alpha_\mathrm{max} > 0$ is a hyperparameter.
When $t\alpha_\mathrm{max}\rightarrow\infty$, the distribution approaches a $\delta$ distribution at either $(1,0)$ or $(0,1)$, as shown in Supporting Figure \ref{fig:conv-dirichlet}.

\begin{figure}
    \centering
    \includegraphics[width=1\linewidth]{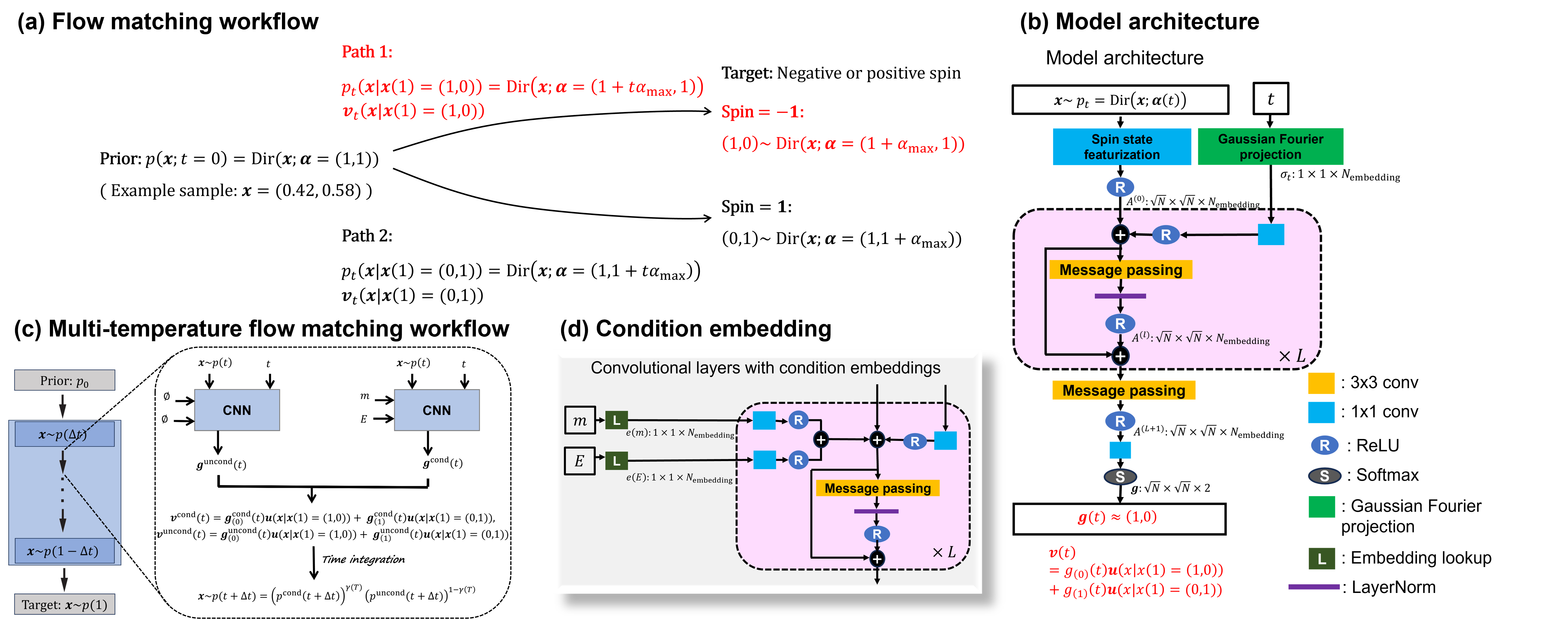}
    \caption{Alchemical flow matching generator. (a) Workflow of the alchemical flow matching model. The spin state of a lattice site is represented as a two-dimensional vector $\boldsymbol{x}(t) = \left( x_{(0)}(t), x_{(1)}(t) \right)$, with $\boldsymbol{x}(1) = (1,0)$ if $s=-1$ and $\boldsymbol{x}(1) = (0,1)$ if $s=1$. The initial state $\boldsymbol{x}(0)$ is sampled from a Dirichlet distribution $\text{Dir}\left(\alpha=(1,1)\right)$, providing uniform random initialization. As $t$ increases, the Dirichlet distribution sharpens toward the target states $(1,0)$ or $(0,1)$.
    (b) A convolutional neural network (CNN) is trained to predict the classifier $\boldsymbol{g}(t)$ that guides the probability flow. The input spin configuration $\boldsymbol{x}$ and time $t$ are separately featurized, then combined through convolutional layers. The lattice size is denoted as $\sqrt{N}\times \sqrt{N}$, and $N$ equals the number of spins in a lattice. The model output $\boldsymbol{g}(t)$ is trained against one-hot labels corresponding to the final spin states. The CNN model is functionally equivalent to a graph neural network. (c) The workflow for multitemperature flow matching by applying guidance facilitated by combining conditional and unconditional generation. The temperature-dependent parameter $\gamma(T)$ controls the temperature of the generated ensemble. (d) Expanded view of the convolutional layers showing incorporation of conditional embeddings for conditioning variables (magnetization $m$ and potential energy $E$). These condition embeddings are added to the feature representations and processed through message-passing blocks.}
    \label{fig:1s1-algo}
\end{figure}

Now that we have chosen a conditional probability path, we design a conditional velocity field that generates this conditional probability path by
\begin{equation}
\boldsymbol{u}_t \left(\boldsymbol{x} \mid \boldsymbol{x}(1)\right) = 
\begin{cases}
     C \left( x_{(0)}, t\alpha_\mathrm{max} \right) \left( \left( 1, 0 \right) - \boldsymbol{x} \right), & \text{if } \boldsymbol{x}(1) = \left( 1, 0 \right) \\
     C \left( x_{(1)}, t\alpha_\mathrm{max} \right) \left( \left(0, 1 \right) - \boldsymbol{x} \right), & \text{if } \boldsymbol{x}(1) = \left( 0, 1 \right).
\end{cases}
\label{eq:u-dirflow}
\end{equation}
The $C \left( x_{(k)}, b \right)$ is derived to be~\cite{stark2024dirichlet}
\begin{equation}
    C \left( x_{(k)}, b \right)=- \frac{\partial}{\partial t} I_{x_{(k)}}\left( b+1, 1 \right)\frac{\mathcal{B} \left( b+1, 1 \right)}{\left( 1-x_{(k)} \right) x_{(k)}^b }
    \label{eq:norm-C-dir}
\end{equation}
where $I_x$ 
is the regularized incomplete beta function, and $\mathcal{B}$ is the multivariate beta function $\mathcal{B} \left( \alpha_0, \alpha_1 \right) = \frac{\Gamma \left( \alpha_0 \right) \Gamma \left( \alpha_1 \right)}{\Gamma \left( \alpha_0 + \alpha_1 \right)}$. It can be proven that the conditional velocity field Eq.~\ref{eq:u-dirflow} and the conditional probability path Eq.~\ref{eq:probpath-dir} together satisfy the transport equation Eq.~\ref{eq:transport} and therefore constitute a valid flow matching framework~\cite{stark2024dirichlet}. Figure~\ref{fig:1s1-algo}(a) illustrates the schematic workflow of Dirichlet flow matching.

Eqs.~\ref{eq:probpath-dir}-\ref{eq:u-dirflow} provide an analytical solution for the variance-exploding path in discrete space. This inherently addresses the challenge of normalizing the probability path, a problem traditionally handled by introducing an explicit normalization term in the loss function and enforcing normalization during training~\cite{campbell2024generative,gat2024discrete,zhao2024probabilistic}.

\subsection{Vector Field Model by CNN}
\label{sec:cnn}

We train a neural network classifier to predict $\boldsymbol{g}(t)$ to approximate $\boldsymbol{x}(1)$ that chooses from the two cases of probability path in Eq.~\ref{eq:probpath-dir}.

\subsubsection{CNN as a Spin Graph}
We use a convolutional neural network (CNN) to predict $\boldsymbol{g}(t)$. This CNN is functionally equivalent to a graph neural network, since convolution can be viewed as a specialized form of message passing over the spin graph. As shown in the model architecture of Figure~\ref{fig:1s1-algo}(b), each Ising configuration is represented as a graph, where spins correspond to nodes and undirected edges connect neighboring spins. 

\subsubsection{Spin State Featurization} 
For each spin $i$, the input $\boldsymbol{x}_i$ is fed into a ReLU-activated $1\times 1$ convolutional layer that outputs a learnable node representation $A_i$ of length $N_\mathrm{embedding}$. $N_\mathrm{embedding}$ is chosen to be $128$. In the following, we use $A_i^{(l)}$ to denote the node representation at message passing layer $l$.

\subsubsection{Time Embedding}
Time is encoded by using a Gaussian Fourier projection
\begin{equation}
    \sigma_t = \left( \sin \left( 2 \pi \omega t \right), \cos \left( 2 \pi \omega t \right) \right)
\end{equation}
where $\omega$ is a learnable weight vector of length $N_\mathrm{embedding}/2$ and the equation outputs a learnable time embedding $\sigma_t$ of length $N_\mathrm{embedding}$.

\subsubsection{Message Passing by Convolutional Layers}
At each message-passing layer, $\sigma_t$ is first mapped to a representation $\mathcal{T}^{(l)}$  of size $N_\mathrm{embedding}$ using a ReLU-activated $1 \times 1$ convolutional layer.
Message passing is then performed using a $3\times 3$ convolutional layer
\begin{equation}
    M_i^{(l)}=\sum_{j \in \mathcal{N}(i) \cup \{i\}} H^{(l)} \left( \vec{r}_{ji} \right) \left( A^{(l)}_{j}+\mathcal{T}^{(l)} \right),
    \label{eq:CNN-message}
\end{equation}
where $H^{(l)}  \left( \vec{r}_{ji} \right)$ consists of nine matrices of dimension $N_\mathrm{embedding}\times N_\mathrm{embedding}$, corresponding to the eight neighbors considered for each spin, along with the self-interaction. 
The node representations are then updated as
\begin{equation}
    A^{(l+1)}_i=\text{ReLU}\left( M_i^{(l)} \right)+A^{(l)}_i.
\end{equation}

\subsubsection{Readout}
After $L=12$ message-passing blocks, the node representations are aggregated using a ReLU-activated $3 \times 3$ convolutional layer, followed by a $1 \times 1$ convolutional layer with $\text{Softmax}$ activation, yielding a two-class classifier $\boldsymbol{g}_i$ for each spin $i$.

\subsection{Loss Functions and Training}
\label{sec:loss}

\subsubsection{Cross Entropy Loss}
Training is conducted via a cross-entropy loss~\cite{stark2024dirichlet}
\begin{equation}
    \mathcal{L}_{\mathrm{CE}}= -\lambda_{\mathrm{CE}}\mathbb{E}_t \sum_i \left[ \boldsymbol{x}_{i}(1)\cdot \ln \boldsymbol{g}_i(t) \right]
    \label{eq:celoss}
\end{equation}
where $\boldsymbol{x}_{i}(1)$ denotes the training data of spin $i$, represented as a one-hot vector taking values $(1,0)$ or $(0,1)$. The prefactor $\lambda_{\mathrm{CE}}$ adjusts the relative weight of the loss term during training.

\subsubsection{Energy-Based Loss}
For the Boltzmann distribution, accurate modeling of the probabilities of configurations with multiple interacting spins can be achieved by training with an energy-based loss function~\cite{noe2019boltzmann,schebek2024efficient,akhound2024iterated}. 

To define the energy-based loss, we first define the energy as a function of spin states $\boldsymbol{x}_i$.
We use $\hat{x}= \{x_{i(k)} \mid i = 1,2,\dots,N;   k = 0,1 \}$ to denote the target state of a configuration with $N$ lattice sites. Similarly, we use $\hat{g}= \{g_{i(k)} \mid i = 1,2,\dots,N;   k = 0,1 \}$ to denote the generated classifiers for a configuration with $N$ lattice sites. To calculate the energy $E$ as a function of $\hat{x}$, we first determine the most likely spin state at each site by
\begin{equation}
s_i = -1 + 2 \ \text{argmax}_{k \in \left\{ 0, 1 \right\}}x_{i(k)} .
\label{eq:s-argmax-x}
\end{equation}
Then we compute the local energy for both spin states: 
$E_i(s_i=-1) = \sum_{j \in \mathcal{N}(i)} s_j
\quad\text{and}\quad
E_i(s_i=1) = -\sum_{j \in \mathcal{N}(i)} s_j$. The total energy of the lattice is obtained as
\begin{equation}
E(\hat{x})=\sum_{\boldsymbol{x}_i \in \hat{x}} \left( \sum_{j \in \mathcal{N}(i)} s_j \right) \left( x_{i(0)} - x_{i(1)} \right)
.
\label{eq:energy-x}
\end{equation}
The generated classifiers $\hat{g}(t)$ and the target states $\hat{x}(1)$ are used to compute the energies $E(\hat{g}(t))$ and $E(\hat{x}(1))$ respectively.

Then, we use the energies to define a loss function
\begin{equation}
    \mathcal{L}_{\mathrm{E}}
    = \lambda_{\mathrm{E}} \mathbb{E}_t \left[
    e^{-E \left( \hat{x}(1) \right) / \tau} \left(E \left( \hat{g}(t) \right) -E \left( \hat{x}(1) \right) \right) /\tau \right] + \lambda_{\mathrm{MAE}} \mathbb{E}_t\left\|E \left( \hat{g}(t) \right) - E \left( \hat{x}(1) \right) \right\|_1
    \label{eq:eloss}
\end{equation}
where $\lambda_{E}$ is the prefactor and $\tau=k_BT$ is the temperature. To avoid numerical instabilities from sharp energy landscapes, training begins at elevated $\tau$ and gradually anneals toward $k_BT$.
Here, we use a mean absolute error term $\mathbb{E} \left\| E \left( \hat{g}(t) \right) - E \left( \hat{x}(1) \right) \right\|_1$ weighted by a prefactor $\lambda_{\mathrm{MAE}}$, since it encourages a tighter alignment of the predicted energy $E \left( \hat{g}(t) \right)$ with the reference energy $E \left( \hat{x}(1) \right)$ and accelerates training.

\subsubsection{Reaction Coordinate Loss}
To help the model distinguish important energy degenerate states, we further use a reaction coordinate loss.
For the square lattice Ising model, the magnetization $m(\hat{x})$ is used as reaction coordinate, which is determined in a similar way as the energy: the most likely spin state is determined by Eq.~\ref{eq:s-argmax-x}, and the local contribution $m_i(s_i=-1)=-1$ and $m_i(s_i=1)=1$ are computed, then the total $m(\hat{x})$ of the lattice is obtained as $m\left( \hat{x} \right) = \sum_{i=1}^N \left( x_{i(1)} - x_{i(0)} \right)$.
Then, the probabilities $P \left( m \left( \hat{x}(1) \right) \right)$ and $P \left( m \left( \hat{g}(t) \right) \right)$ are computed by batchwise kernel density estimation over the reaction coordinates of the training samples and the predicted classifiers respectively~\cite{noe2019boltzmann}.
The reaction coordinate loss is defined as
\begin{equation}
\mathcal{L}_{\mathrm{RC}} = \lambda_{\mathrm{RC}} \mathbb{E}_t D_{\mathrm{KL}} \left[ P \left( m \left( \hat{x}(1) \right) \right) \Vert P \left( m \left( \hat{g}(t) \right) \right) \right]
\label{eq:rcloss}
\end{equation}
where $\lambda_{\mathrm{RC}}$ is the prefactor.

\subsubsection{Training}
Models are trained with the Adam optimizer~\cite{kingma2014adam} using an initial learning rate of $5\times10^{-4}$ over batch sizes of 1,024 lattice configurations. Details on the scheduling of prefactors for different loss functions are provided in the Supporting Section~\ref{app:training-cost}.

\subsection{Flow Matching Inference}
\label{sec:FM-inference}
At inference, we parameterize the marginal vector field using the model prediction $\boldsymbol{g}_i(t)$ via~\cite{stark2024dirichlet}
\begin{equation}
    \boldsymbol{v}_{i}(t)
    =  g_{i(0)}(t) \boldsymbol{u}_t \left( \boldsymbol{x}_i \mid \boldsymbol{x}_i(1) = (1,0) \right) + g_{i(1)}(t) \boldsymbol{u}_t \left( \boldsymbol{x}_i \mid \boldsymbol{x}_i(1) = (0,1) \right)
    \label{eq:v-dirflow}
\end{equation}
where $\boldsymbol{u}_t \left( \boldsymbol{x} \mid \boldsymbol{x}(1) \right)$ is given in Eq.~\ref{eq:u-dirflow}. 

The marginal velocity is integrated over time to trace the marginal probability path by 
\begin{equation}
    \boldsymbol{x}_{i}(1)-\boldsymbol{x}_{i}(0)=\int_0^1 \boldsymbol{v}_{i}(t) \mathrm{d}t.
\end{equation}
In practice, the probability path converges to the target distribution $\delta(\boldsymbol{x}_i-\boldsymbol{x}_{i}(1))$ at $t\alpha_\mathrm{max} \ge 9$, as shown in Supporting Figure~\ref{fig:conv-dirichlet}, so we set $\alpha_\mathrm{max}=9$. We employed 80 integration steps, as additional steps yield essentially the same results.

\section{Reproducing the Free Energy Surface of the Ising Model}\label{sect:fes-ising-uncond}

We first apply alchemicalFES to generate the FES of the 2D square-lattice Ising model. 
The training set was generated using MCMC sampling of a $6 \times 6$ lattice Ising model at a given target temperature. Since the goal is to train the model to accurately capture the statistics of the dataset, the dataset must be large enough to contain a reasonable number of low-probability samples. In this work, we use a dataset with 1 million samples. 

The trained model was then utilized to generate samples for larger lattice sizes. Ensemble averaging was then used to construct the FES in an order parameter space. Specifically, we used two order parameters: the magnetization per spin ($m/N = \frac{1}{N} \sum_{i=1}^{N} s_{i}$) and the potential energy per spin ($E/N = -\frac{1}{N}\frac{1}{2} \sum_{i=1}^N \sum_{j=1}^N J_{ij} s_i s_j$).

\begin{figure}[!htp]
    \centering
    \includegraphics[width=1\linewidth]{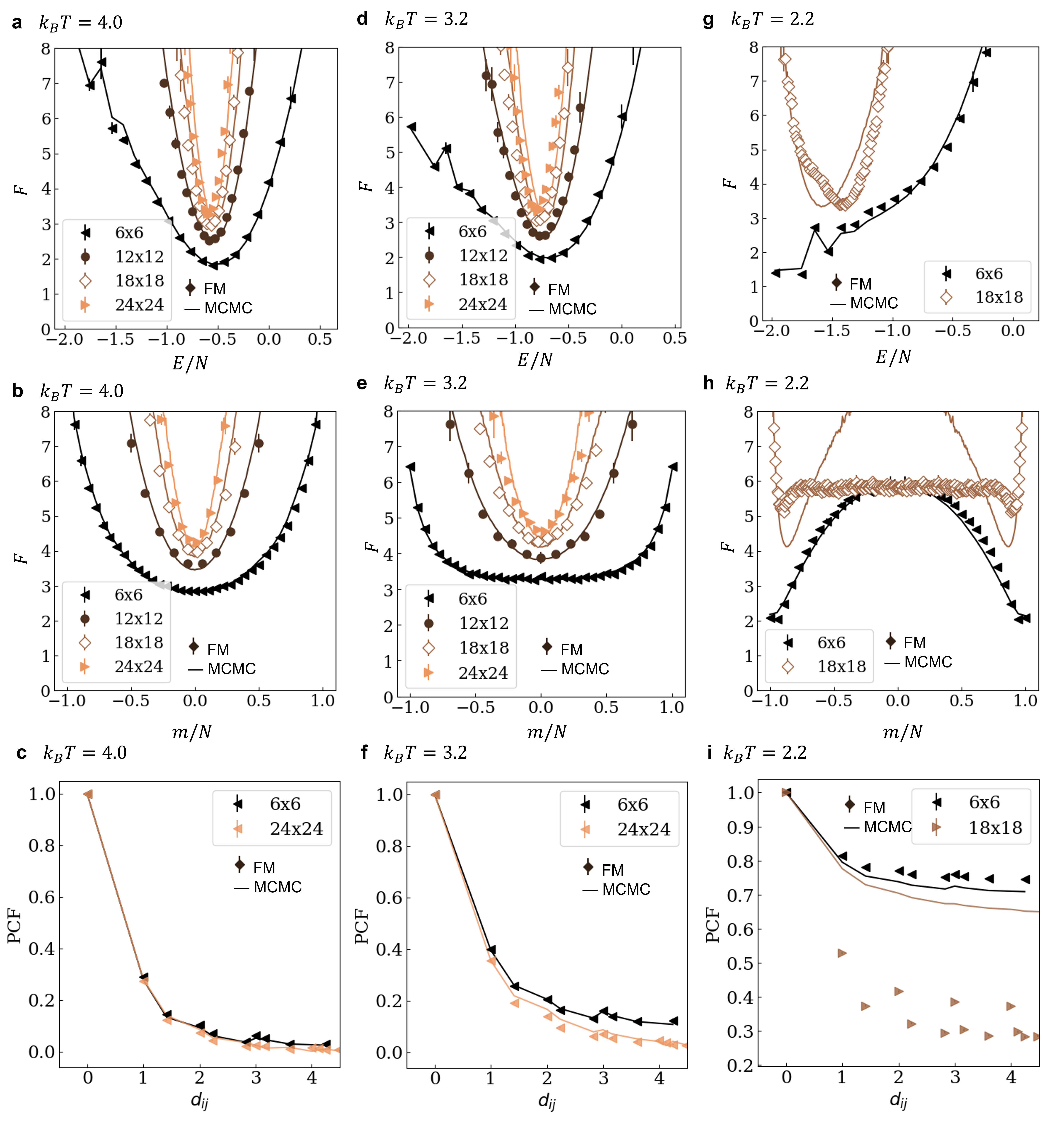}
    \caption{Free energy estimations from flow matching model (dots) for Ising lattice of different sizes, compared against the free energy surface from MCMC simulations (lines). Free energy as a function of $E/N=-\frac{1}{N} \frac{1}{2}\sum_{i=1}^N \sum_{j=1}^N J_{ij} s_is_j$ at (a) $k_BT=4.0$, (d) $k_BT=3.2$, (g) $k_BT=2.2$; free energy as a function of $m/N=\frac{1}{N}\sum_{i=1}^N s_i$ at (b) $k_BT=4.0$, (e) $k_BT=3.2$, (h) $k_BT=2.2$; pair correlation function (PCF) at (c) $k_BT=4.0$, (f) $k_BT=3.2$, (i) $k_BT=2.2$. 
    There is a kink at $E/N\approx -1.667$ due to the finite-size effect of the Ising model. We explain this peculiar phenomenon in Supporting Section~\ref{app:finite-size-effect}.}
    \label{fig:2s1-uncond}
\end{figure}

The results are shown in Figure~\ref{fig:2s1-uncond}.
For each lattice site, only eight neighbors are considered. As a result, the model prediction is size scalable under the limit of short-range correlations that rapidly diminish with distance. This assumption is valid at high temperatures $T>T_c$, where thermal fluctuations disrupt long-range order, as shown in Figure~\ref{fig:2s1-uncond}(a,b,d,e). The predicted FES of $k_BT=4.0$ and $k_BT=3.2$ closely match the reference FES for various lattice sizes. Figure~\ref{fig:2s1-uncond}(c,f) shows the pair correlation function (PCF). For both $6\times6$ lattices and $24\times24$ lattices, the predicted PCF matches the references very well.  
However, for low temperatures, the system exhibits long-range correlation and the naive flow matching approach no longer maintains size scalability, as shown in Figure~\ref{fig:2s1-uncond}(h,i). 

\subsection{Size Scalable Multitemperature Generation}\label{sect:multi-T-FM}

In this section, we demonstrate how to generate samples across multiple temperatures with a single model by leveraging the \textit{classifier-free guidance} technique and how this approach leads to size scalability even at low temperatures with long-range correlation. In the following, we briefly review the guidance technique and then describe how we adapted it to enable size-scalable generation across multiple temperatures.

\subsubsection{Guidance Technique}
A key feature of iterative generative models is their ability to progressively bias the generative process toward specific target distributions, a concept known as \textit{guidance}~\cite{dhariwal2021diffusion,ho2022classifier}. Guidance was originally introduced in the context of diffusion models to learn the score $\boldsymbol{\epsilon}(\boldsymbol{x}; t) = \nabla_{\boldsymbol{x}} \ln p_t(\boldsymbol{x})$ of the distribution of noisy data. In particular, \textit{classifier guidance} modifies the score by incorporating the gradient of an auxiliary classifier's log-likelihood~\cite{dhariwal2021diffusion}
\begin{equation}
    \boldsymbol{\epsilon} \left( \boldsymbol{x}, c; t \right) = \nabla_{\boldsymbol{x}} \ln p_t \left( \boldsymbol{x} \right) + \gamma  \nabla_{\boldsymbol{x}} \ln p \left( c \mid \boldsymbol{x} \right)
    \label{eq:cg}
\end{equation}
where $c$ denotes the desired class, and $\gamma$ controls the strength of the classifier guidance.

To remove the need for a separate classifier model, Ho and Salimans~\cite{ho2022classifier} introduced \textit{classifier-free guidance}, which substitutes
$p \left(c \mid \boldsymbol{x} \right) = p_t \left( \boldsymbol{x} \mid c \right) p_t \left( c \right) / p_t \left( \boldsymbol{x} \right)$ into Eq.~\eqref{eq:cg}, leading to 
\begin{equation}
    \boldsymbol{\epsilon} \left( \boldsymbol{x}, c; t \right) = \gamma \nabla_{\boldsymbol{x}} \ln p_t \left( \boldsymbol{x} \mid c \right) + \left( 1 - \gamma \right) \nabla_{\boldsymbol{x}} \ln p_t \left( \boldsymbol{x} \right)
\end{equation}
where we train a conditional model to learn $\nabla_{\boldsymbol{x}} \ln p_t \left( \boldsymbol{x} \mid c \right)$.

\subsubsection{Relationship between Flow and Score}
For the Dirichlet probability path, the score can be obtained from the model posterior via the denoising score-matching identity~\cite{song2019generative}
\begin{equation}
    \boldsymbol{\epsilon}_t(\boldsymbol{x}) = \boldsymbol{\epsilon}_t \left( \boldsymbol{x} \mid \boldsymbol{x}(1) = (1,0) \right) p_t \left( \boldsymbol{x} \mid \boldsymbol{x}(1) = (1,0) \right) + \boldsymbol{\epsilon}_t \left( \boldsymbol{x} \mid \boldsymbol{x}(1) = (0,1) \right) p_t \left( \boldsymbol{x} \mid \boldsymbol{x}(1) = (0,1) \right)
    \label{eq:marginal-score}
\end{equation}
where $p_t \left( \boldsymbol{x} \mid \boldsymbol{x}(1) \right)$ is given in Eq.~\ref{eq:probpath-dir}.
Let $\boldsymbol{p} = \left( p_t \left( \boldsymbol{x} \mid \boldsymbol{x}(1) = (1,0) \right), p_t \left( \boldsymbol{x} \mid \boldsymbol{x}(1) = (0,1) \right) \right)$ be a vector containing the two cases of Eq.~\ref{eq:probpath-dir}.
We can differentiate $\boldsymbol{p}$ w.r.t. $\boldsymbol{x}$ to obtain a $2 \times 2$ Jacobian matrix
$\mathrm{diag} \left( \boldsymbol{\epsilon} \right) = D \mathrm{\ diag} \left( \boldsymbol{p} \right)$
where $\boldsymbol{\epsilon} \in \mathbb{R}^2$ and $D = \mathrm{diag} \left( t / \boldsymbol{x} \right) \in \mathbb{R}^{2 \times 2}$.
We can rewrite the Jacobian equation to
$\boldsymbol{\epsilon} = D \boldsymbol{p}$.
Meanwhile, the computation of the marginal flow (Eq.~\ref{eq:v-dirflow}) can also be written as a very similar matrix equation $\boldsymbol{v} = U \boldsymbol{g}$ where $\boldsymbol{g} = \left( g_{(0)}(t), g_{(1)}(t) \right)$ and the entries of $U$ are given by Eq.~\ref{eq:u-dirflow}:
\begin{equation}
U = \left[ \begin{matrix}
\boldsymbol{u}_t \left( \boldsymbol{x} \mid \boldsymbol{x}(1) = (1,0) \right) \\
\boldsymbol{u}_t \left( \boldsymbol{x} \mid \boldsymbol{x}(1) = (0,1) \right)
\end{matrix} \right]^T \in \mathbb{R}^{2 \times 2} .
\end{equation}
Using $\boldsymbol{g} \approx \boldsymbol{p}$, we obtain
\begin{equation}
\boldsymbol{v} = U \boldsymbol{g} = U \boldsymbol{p} = U D^{-1}\boldsymbol{\epsilon} .
\label{eq:score-to-flow}
\end{equation}
Thus, a linear relationship exists between the marginal flow $\boldsymbol{v}$ and the score $\boldsymbol{\epsilon}$ arising from the model posterior $\boldsymbol{p}$. 

Suppose we have conditional and unconditional flow models $\boldsymbol{v}\left( \boldsymbol{x}; t \mid c \right)$ and $\boldsymbol{v} \left( \boldsymbol{x}; t \right)$. Since a linear combination of scores results in a linear combination of flows, we similarly implement guidance to the flows by integrating
\begin{equation}
\boldsymbol{v}_\mathrm{CFG} \left( \boldsymbol{x}, c; t \right) = \gamma \boldsymbol{v} \left( \boldsymbol{x}; t \mid c \right) + \left( 1 - \gamma \right) \boldsymbol{v} \left( \boldsymbol{x}; t \right)
.    
\end{equation}

\subsubsection{Multitemperature Generation via Classifier-Free Guidance}

A straightforward way to enable multitemperature generation with a single model is to treat \textit{temperature} as a conditioning variable. Schebek et al.~\cite{schebek2024efficient} demonstrated a realization of this by using conditional normalizing flows, where temperature and pressure were used as input features to predict free energy differences between solid and liquid phases. However, they noted that this approach required more advanced model architectures and much longer training times compared to their earlier work, which addressed prediction under a single thermal condition.  Moreover, extending it to reproduce the entire FES under multiple conditions would demand even greater training effort.

In this work, we propose a two-step approach to accomplish multitemperature generation without increasing model capacity or incurring substantially greater training costs.

\paragraph{Step 1: Using Order Parameters as Conditioning Variables.}

The first idea is to include more information in the conditioning variables. Rather than conditioning only on the temperature, we condition on the order parameters of a Ising model at the given temperature. 
For the Ising model defined by the Hamiltonian $H = -\frac{1}{2}\sum_{i=1}^N \sum_{j=1}^N J_{ij} s_i s_j$, we condition our flow model on two order parameters: the magnetization $m = \sum_{i=1}^N s_i$ and the energy $E = H$.

Figure~\ref{fig:1s1-algo}(d) illustrates the modified convolutional layers incorporating condition embeddings. Since the energy $E$ and magnetization $m$ of a $6\times 6$ lattice Ising model take discrete values, we employed an embedding lookup of length $N_\mathrm{embedding}$ to represent the $E$ and $m$ conditions as learnable embeddings $e(E)$ and $e(m)$. The index set for $m$ is $\left\{ n \mid -36 \leq n \leq 36, n \equiv 0 \pmod{2} \right\}$, and the index set for $E$ is $\left\{ n \mid -72 \leq n \leq 72, n \equiv 0 \pmod{4} \right\}$. For larger lattices, we rescale $E$ and $m$ by $N/36$ to map them to the lookup range, where $N$ denotes the number of lattice sites. Then, a ReLU-activated linear layer maps $e(E)$ and $e(m)$ to energy and magnetization features of length $N_\mathrm{embedding}$ at message passing layer $l$: $\mathcal{E}^{(l)}$ and $\mathcal{M}^{(l)}$. The message passing is augmented with energy and magnetization features, yielding
\begin{equation}
    M_i^{(l)} = \sum_{j \in \mathcal{N}(i)\cup \{i\}} H^{(l)} \left( \vec{r}_{ji} \right) \left( A^{(l)}_{j} + \mathcal{T}^{(l)} + \mathcal{E}^{(l)} + \mathcal{M}^{(l)} \right).
\end{equation}

\begin{figure}
    \centering
    \includegraphics[width=1\linewidth]{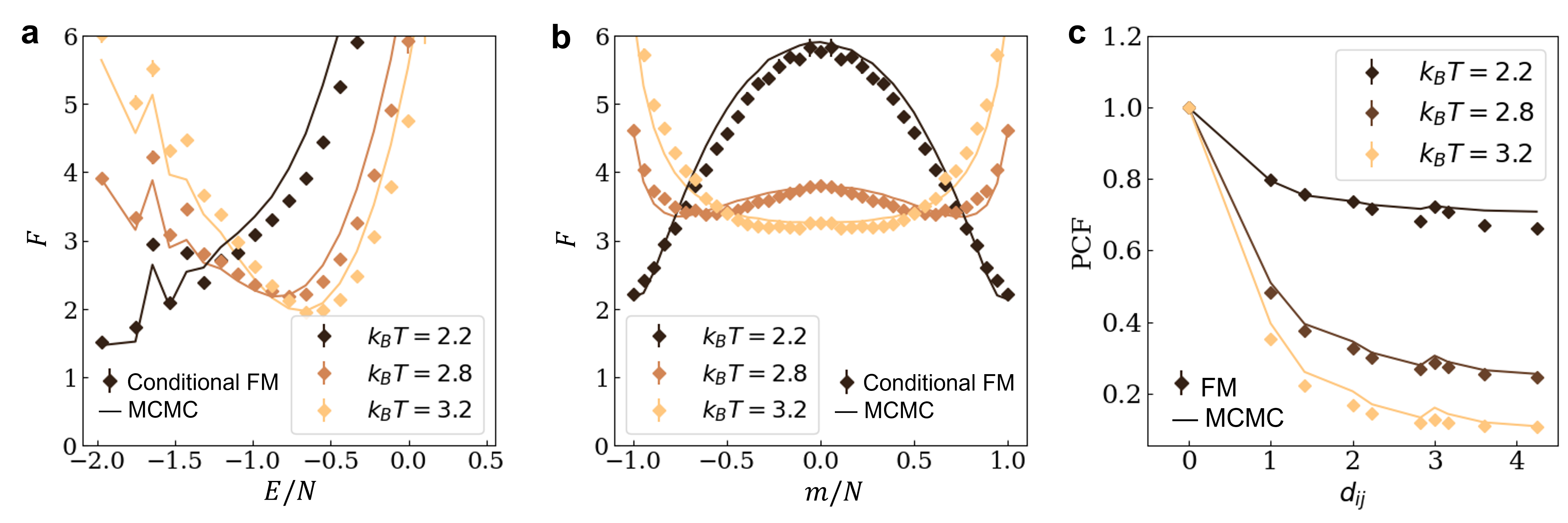}
    \caption{Conditional generation (dots) of $6 \times 6$ lattice Ising model using order parameter conditions, compared against MCMC data (lines). (a) Free energy as a function of $E/N=-\frac{1}{N}\frac{1}{2}\sum_{i=1}^N \sum_{j=1}^N J_{ij} s_is_j$; (b) free energy as a function of $m/N=\frac{1}{N}\sum_{i=1}^N s_i$; and (c) pair correlation function (PCF).}
    \label{fig:1s3-cond-6x6}
\end{figure}

The conditional model was trained on the data of $k_BT=3.2,\ 2.8,\ 2.4,\ 2.2,\ 2.0,\ 0.0$. Figure~\ref{fig:1s3-cond-6x6} shows the FES generated for a $6 \times 6$ lattice under this scheme.

\paragraph{Step 2: Size-Scalable Generation with Temperature-Dependent Guidance.}

However, conditional generation with order parameter conditions alone cannot scale to larger system sizes, because we lack the order parameters for bigger lattices. To address scalability, we turn to the guidance technique. First, we define the guided score of the spin $i$ as
\begin{equation}
\begin{split}
\boldsymbol{\epsilon}_{\mathrm{CFG},i} \left( \hat{x}, E, m ; t \right)
&= \gamma \nabla_{\boldsymbol{x}_i} \ln P \left( \hat{x}; t \mid E, m \right) + \left( 1 - \gamma \right) \nabla_{\boldsymbol{x}_i} \ln P \left( \hat{x}; t \right) \\
&= \nabla_{\boldsymbol{x}_i} \ln \left[ P \left( \hat{x}; t \mid E, m \right)^{\gamma} P \left( \hat{x}; t \right)^{1-\gamma} \right]
\end{split}
\label{eq:s-cfg}
\end{equation}
where $P(\hat{x},t)$ denotes the predicted probabilities of sample $\hat{x}$ at time $t$. This implies the guided distribution 
\begin{equation}
    P_{\mathrm{CFG}} \left( \hat{x}, E, m ; t \right) = \frac{P \left( \hat{x}; t \mid E, m \right)^{\gamma} P \left( \hat{x}; t \right)^{1-\gamma}}{Z_{\mathrm{CFG}} \left( t, E, m \right)}
    \label{eq:p-cfg}
\end{equation}
where $Z_{\mathrm{CFG}} \left( t, E, m \right)$ is a normalizing constant. 

The Boltzmann probability of a sample $\hat{x}$ at a target temperature $T$ is
\begin{equation}
    P_T \left( \hat{x} \right) = \frac{1}{Z_T} \exp \left( -\frac{E \left( \hat{x} \right)}{k_B T} \right)
    \label{eq:p_T}
\end{equation}
where $Z_T$ is a normalizing constant. The score of spin $i$ is 
\begin{equation}
\nabla_{\boldsymbol{x}_i} \ln P_T \left( \hat{x} \right) = -\frac{1}{k_BT}\nabla_{\boldsymbol{x}_i} E \left( \hat{x} \right).
\label{eq:s_T}
\end{equation}
We denote the temperature of the ensemble generated by conditional FM by $T^{\mathrm{cond}}$, and that generated by unconditional FM by $T^{\mathrm{uncond}}$. Their respective scores at $t=1$ are 
$\frac{1}{k_BT^{\mathrm{cond}}} \nabla_{\boldsymbol{x}_i} E \left( \hat{x}(1) \right)$
and $\frac{1}{k_BT^{\mathrm{uncond}}} \nabla_{\boldsymbol{x}_i} E \left( \hat{x}(1) \right)$ respectively.
Then, the guided score Eq.~\eqref{eq:s-cfg} at $t=1$ can be written as
\begin{equation}
    \boldsymbol{\epsilon}_{\mathrm{CFG},i}\left( \hat{x}, E, m; 1 \right)
    =-\frac{\gamma}{k_BT^{\mathrm{cond}}}\nabla_{\boldsymbol{x}_i} E \left( \hat{x}(1) \right) - \frac{1-\gamma}{k_BT^{\mathrm{uncond}}}\nabla_{\boldsymbol{x}_i}E \left( \hat{x}(1) \right).
    \label{eq:epsilon-cfg}
\end{equation}
$\boldsymbol{\epsilon}_{\mathrm{CFG},i}\left( \hat{x}, E, m ; 1 \right)$ coincides with the score of the Boltzmann distribution at the target temperature $T$ if the guidance parameter $\gamma$ is chosen to satisfy
\begin{equation}
    \frac{1}{k_B T} = \frac{\gamma}{k_B T^{\mathrm{cond}}} + \frac{1-\gamma}{k_B T^{\mathrm{uncond}}} ,
    \label{eq:temperature-bias}
\end{equation}
i.e.,
\begin{equation}
\gamma = \frac{T^{\mathrm{cond}} \left( T^{\mathrm{uncond}} - T \right)}{T \left( T^{\mathrm{uncond}} - T^{\mathrm{cond}} \right)} ,
\end{equation}
thus ensuring that $\boldsymbol{\epsilon}_{\mathrm{CFG},i}\left( \hat{x}, E, m; 1 \right)$ reproduces the correct Boltzmann distribution at temperature $T$. 
Furthermore, because the Dirichlet flow follows a linear relationship with the score, we can obtain the corresponding flow using the guided score by Eq.~\ref{eq:score-to-flow}. The multitemperature generation algorithm is depicted in Figure~\ref{fig:1s1-algo}(c).

Size scalability is maintained even at low temperatures under this scheme. When generating configurations at multiple temperatures for lattice sizes larger than those in the training set, we use as input conditions the order parameters of the magnetic states (i.e., all spins up or all spins down, whose magnetizations $m$ and energies $E$ are readily calculated), which leads to the generation of a low-temperature ensemble. The desired temperature can then be obtained by tuning the $\gamma$ parameter in Eq.~\ref{eq:epsilon-cfg} according to Eq.~\ref{eq:temperature-bias}.

The \emph{classifier-free guidance} technique here allows us to reuse the same flow model to generate for multiple temperatures with minimal architectural changes or additional training effort.

\subsubsection{Reproducing the Free Energy Surface of the Ising Model at Multiple Temperatures}

\begin{figure}
    \centering
    \includegraphics[width=1\linewidth]{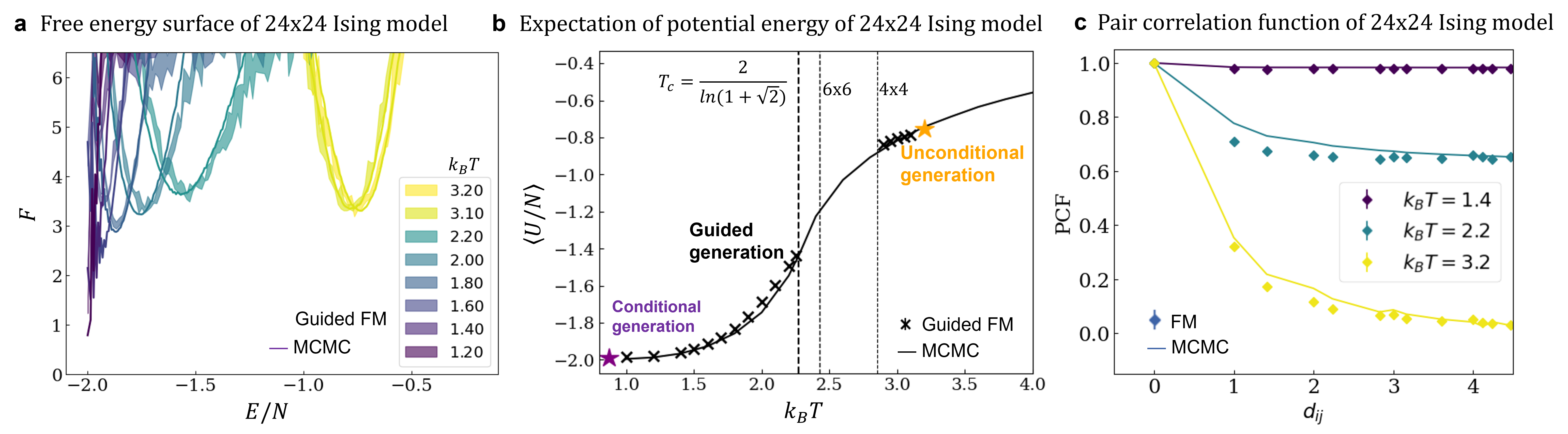}
    \caption{(a) Free energy estimations for $24 \times 24$ lattice Ising model at multiple temperatures obtained by the guided FM model, trained with the MCMC data of $6\times6$ lattice Ising model, are compared against the MCMC free energies (lines). The shaded region illustrates the 97.5\% confidence interval of the estimated free energy. (b) Potential energy expectations at multiple temperatures predicted by the guided FM model (crosses), compared against MCMC data (lines). The vertical lines indicate the critical temperatures of phase transition for different lattice sizes, where $T_c=\frac{2}{\ln (1+\sqrt{2})}$ in the infinite lattice limit~\cite{binder1981finite}. $T_c\approx 2.43$ and $T_c\approx 2.85$ are estimated for $6\times 6$ and $4\times 4$ Ising lattice respectively~\cite{kadanoff1966scaling}, by the renormalization group theory~\cite{wilson1971renormalization1,wilson1971renormalization2,wilson1983renormalization}. (c) Pair correlation function of multiple temperatures obtained by the guided FM model, compared against the MCMC data.}
    \label{fig:2s2-guided}
\end{figure}

Using the magnetic states as input conditions, the exact value of $k_BT^{\mathrm{cond}}$ is determined by fitting the generated FES of a $6\times 6$ lattice via Eq.~\eqref{eq:p-cfg} to a reference FES of the training data. Specifically, we iteratively adjust $\gamma$ until the generated FES best matches the reference. Substituting the resulting $\gamma$ back into Eq.~\eqref{eq:p-cfg} yields the desired $k_B T^{\mathrm{cond}}$. By fitting the FES at $k_BT=2.2$, we obtained $k_BT^{\mathrm{cond}}\approx 0.872$. The corresponding $\gamma(T)$ values are shown in Supporting Figure~\ref{fig:gamma-T}. Next, by applying Eq.~\ref{eq:p-cfg} using the conditional model and the unconditional flow model of $k_BT=3.2$, we were able to generate the full temperature range within $0.872 < k_B T < 3.2$ for a large $24\times 24$ Ising model as shown in Figure~\ref{fig:2s2-guided}, except for a range near the critical temperature of phase transition. 
Near the critical temperature, a phase transition between ordered and disordered states takes place and the gradient of the free energy diverges, rendering Eq.~\ref{eq:s-cfg} invalid in that regime.

\section{Discussion}
In this work, we have introduced alchemicalFES, a flow matching model designed for free energy sampling in the discrete alchemical spaces and applied it to the 2D square-lattice Ising model. 
A notable strength of the proposed method lies in its ability to generate the free energy surface at multiple temperatures. We achieve this by adopting classifier-free guidance-based strategies: first, we train a conditional flow model using temperature-dependent order parameters as conditions; second, we employ a reweighting scheme that combines the conditional and unconditional models. By changing a single guidance parameter, we can accurately reproduce the free energy surfaces across a broad temperature range, including both higher temperatures, characteristic of short-range correlations, and lower temperatures with long-range correlations. In addition to its relevance to chemistry and materials science, the study represents an early attempt to employ the guidance technique, which has so far been used primarily for qualitative control, to achieve quantitative control over the probability distributions generated by a generative model.

Despite these encouraging results, some limitations persist. Although the method can be readily adapted to other discrete systems, such as solid-state compounds, its effectiveness in three-dimensional settings has not yet been validated, and system-specific challenges inherent to higher-dimensional or more complex configurations warrant further investigation.

\section{Conclusion}
We present a discrete flow matching model, alchemicalFES, that maps a uniform distribution to the target Boltzmann distribution of a 2D Ising spin model, overcoming many of the limitations inherent to both MCMC and the current generative models. Through the introduction of classifier-free guidance-based techniques, we demonstrated the feasibility of a single flow matching model capable of generating the free energy surfaces at multiple temperatures and lattice sizes with minimal training overhead. Our numerical results on the 2D Ising model verify the scalability and accuracy of the approach. Future research directions include extending the method to more complex systems, such as the alchemical space of crystalline compounds.

\section*{Associated Content}
\addcontentsline{toc}{section}{Associated Content}

\subsection*{Supporting Information}

Additional training details, algorithms, analyses, and figures as mentioned in the text.

\section*{Author Notes}
\addcontentsline{toc}{section}{Author Notes}

The authors declare the following competing financial interest(s):
B.C. has an equity stake in AIMATX Inc.

\section*{Acknowledgments}
\addcontentsline{toc}{section}{Acknowledgments}

P.T. acknowledges funding from FFG MAGNIFICO and the BIDMaP Postdoctoral Fellowship. Z.Z. acknowledges funding from the European Union's Horizon 2020 research and innovation program under the Marie Skłodowska-Curie grant agreement No. 101034413.
The authors acknowledge the research computing facilities provided by the Institute of Science and Technology Austria (ISTA), and resources of the National Energy Research Scientific Computing Center (NERSC), a Department of Energy Office of Science User Facility using NERSC award DOEERCAP0031751 `GenAI@NERSC'.
P.T. acknowledges valued discussions with Dr. Daniel King, Dr. Lei Wang, and Dr. Fuzhi Dai.

\bibliographystyle{unsrt}
\bibliography{references}

@article{stark2024dirichlet,
  title={Dirichlet flow matching with applications to dna sequence design},
  author={Stark, Hannes and Jing, Bowen and Wang, Chenyu and Corso, Gabriele and Berger, Bonnie and Barzilay, Regina and Jaakkola, Tommi},
  journal={arXiv preprint arXiv:2402.05841},
  year={2024}
}

@article{song2019generative,
  title={Generative modeling by estimating gradients of the data distribution},
  author={Song, Yang and Ermon, Stefano},
  journal={Advances in neural information processing systems},
  volume={32},
  year={2019}
}

@article{binder1981finite,
  title={Finite size scaling analysis of Ising model block distribution functions},
  author={Binder, Kurt},
  journal={Zeitschrift f{\"u}r Physik B Condensed Matter},
  volume={43},
  pages={119--140},
  year={1981},
  publisher={Springer}
}

@article{noe2019boltzmann,
  title={Boltzmann generators: Sampling equilibrium states of many-body systems with deep learning},
  author={No{\'e}, Frank and Olsson, Simon and K{\"o}hler, Jonas and Wu, Hao},
  journal={Science},
  volume={365},
  number={6457},
  pages={eaaw1147},
  year={2019},
  publisher={American Association for the Advancement of Science}
}

@article{hukushima1996exchange,
  title={Exchange Monte Carlo method and application to spin glass simulations},
  author={Hukushima, Koji and Nemoto, Koji},
  journal={Journal of the Physical Society of Japan},
  volume={65},
  number={6},
  pages={1604--1608},
  year={1996},
  publisher={The Physical Society of Japan}
}

@article{kirkpatrick1983optimization,
  title={Optimization by simulated annealing},
  author={Kirkpatrick, Scott and Gelatt Jr, C Daniel and Vecchi, Mario P},
  journal={science},
  volume={220},
  number={4598},
  pages={671--680},
  year={1983},
  publisher={American association for the advancement of science}
}

@article{marinari1992simulated,
  title={Simulated tempering: a new Monte Carlo scheme},
  author={Marinari, Enzo and Parisi, Giorgio},
  journal={Europhysics letters},
  volume={19},
  number={6},
  pages={451},
  year={1992},
  publisher={IOP Publishing}
}

@book{frenkel2023understanding,
  title={Understanding molecular simulation: from algorithms to applications},
  author={Frenkel, Daan and Smit, Berend},
  year={2023},
  publisher={Elsevier}
}

@article{hastings1970monte,
  title={Monte Carlo sampling methods using Markov chains and their applications},
  author={Hastings, W Keith},
  journal={Biometrika},
  volume={57},
  year={1970},
  publisher={Oxford University Press}
}

@article{klein2024equivariant,
  title={Equivariant flow matching},
  author={Klein, Leon and Kr{\"a}mer, Andreas and No{\'e}, Frank},
  journal={Advances in Neural Information Processing Systems},
  volume={36},
  year={2024}
}

@article{schebek2024efficient,
  title={Efficient mapping of phase diagrams with conditional normalizing flows},
  author={Schebek, Maximilian and Invernizzi, Michele and No{\'e}, Frank and Rogal, Jutta},
  journal={arXiv preprint arXiv:2406.12378},
  year={2024}
}

@article{moqvist2024thermodynamic,
  title={Thermodynamic Interpolation: A generative approach to molecular thermodynamics and kinetics},
  author={Moqvist, Selma and Chen, Weilong and Schreiner, Mathias and N{\"u}ske, Feliks and Olsson, Simon},
  journal={arXiv preprint arXiv:2411.10075},
  year={2024}
}

@article{causer2024discrete,
  title={Discrete generative diffusion models without stochastic differential equations: a tensor network approach},
  author={Causer, Luke and Rotskoff, Grant M and Garrahan, Juan P},
  journal={arXiv preprint arXiv:2407.11133},
  year={2024}
}

@article{lou2023discrete,
  title={Discrete diffusion language modeling by estimating the ratios of the data distribution},
  author={Lou, Aaron and Meng, Chenlin and Ermon, Stefano},
  journal={arXiv preprint arXiv:2310.16834},
  year={2023}
}

@article{ghazvininejad2019mask,
  title={Mask-predict: Parallel decoding of conditional masked language models},
  author={Ghazvininejad, Marjan and Levy, Omer and Liu, Yinhan and Zettlemoyer, Luke},
  journal={arXiv preprint arXiv:1904.09324},
  year={2019}
}

@inproceedings{chang2022maskgit,
  title={Maskgit: Masked generative image transformer},
  author={Chang, Huiwen and Zhang, Han and Jiang, Lu and Liu, Ce and Freeman, William T},
  booktitle={Proceedings of the IEEE/CVF Conference on Computer Vision and Pattern Recognition},
  pages={11315--11325},
  year={2022}
}

@article{wu2019solving,
  title={Solving statistical mechanics using variational autoregressive networks},
  author={Wu, Dian and Wang, Lei and Zhang, Pan},
  journal={Physical review letters},
  volume={122},
  number={8},
  pages={080602},
  year={2019},
  publisher={APS}
}

@article{sharir2020deep,
  title={Deep autoregressive models for the efficient variational simulation of many-body quantum systems},
  author={Sharir, Or and Levine, Yoav and Wies, Noam and Carleo, Giuseppe and Shashua, Amnon},
  journal={Physical review letters},
  volume={124},
  number={2},
  pages={020503},
  year={2020},
  publisher={APS}
}

@article{han2022ssd,
  title={Ssd-lm: Semi-autoregressive simplex-based diffusion language model for text generation and modular control},
  author={Han, Xiaochuang and Kumar, Sachin and Tsvetkov, Yulia},
  journal={arXiv preprint arXiv:2210.17432},
  year={2022}
}

@article{olehnovics2024assessing,
  title={Assessing the accuracy and efficiency of free energy differences obtained from reweighted flow-based probabilistic generative models},
  author={Olehnovics, Edgar and Liu, Yifei Michelle and Mehio, Nada and Sheikh, Ahmad Y and Shirts, Michael R and Salvalaglio, Matteo},
  journal={Journal of Chemical Theory and Computation},
  volume={20},
  number={14},
  pages={5913--5922},
  year={2024},
  publisher={ACS Publications}
}

@inproceedings{avdeyev2023dirichlet,
  title={Dirichlet diffusion score model for biological sequence generation},
  author={Avdeyev, Pavel and Shi, Chenlai and Tan, Yuhao and Dudnyk, Kseniia and Zhou, Jian},
  booktitle={International Conference on Machine Learning},
  pages={1276--1301},
  year={2023},
  organization={PMLR}
}

@book{kotz2019continuous,
  title={Continuous multivariate distributions, Volume 1: Models and applications},
  author={Kotz, Samuel and Balakrishnan, Narayanaswamy and Johnson, Norman L},
  volume={334},
  year={2019},
  publisher={John Wiley \& Sons}
}

@article{austin2021structured,
  title={Structured denoising diffusion models in discrete state-spaces},
  author={Austin, Jacob and Johnson, Daniel D and Ho, Jonathan and Tarlow, Daniel and Van Den Berg, Rianne},
  journal={Advances in Neural Information Processing Systems},
  volume={34},
  pages={17981--17993},
  year={2021}
}

@article{hoogeboom2021argmax,
  title={Argmax flows and multinomial diffusion: Learning categorical distributions},
  author={Hoogeboom, Emiel and Nielsen, Didrik and Jaini, Priyank and Forr{\'e}, Patrick and Welling, Max},
  journal={Advances in Neural Information Processing Systems},
  volume={34},
  pages={12454--12465},
  year={2021}
}

@article{gat2024discrete,
  title={Discrete Flow Matching},
  author={Gat, Itai and Remez, Tal and Shaul, Neta and Kreuk, Felix and Chen, Ricky TQ and Synnaeve, Gabriel and Adi, Yossi and Lipman, Yaron},
  journal={arXiv preprint arXiv:2407.15595},
  year={2024}
}

@article{campbell2024generative,
  title={Generative flows on discrete state-spaces: Enabling multimodal flows with applications to protein co-design},
  author={Campbell, Andrew and Yim, Jason and Barzilay, Regina and Rainforth, Tom and Jaakkola, Tommi},
  journal={arXiv preprint arXiv:2402.04997},
  year={2024}
}

@article{campbell2024trans,
  title={Trans-dimensional generative modeling via jump diffusion models},
  author={Campbell, Andrew and Harvey, William and Weilbach, Christian and De Bortoli, Valentin and Rainforth, Thomas and Doucet, Arnaud},
  journal={Advances in Neural Information Processing Systems},
  volume={36},
  year={2024}
}

@article{dhariwal2021diffusion,
  title={Diffusion models beat gans on image synthesis},
  author={Dhariwal, Prafulla and Nichol, Alexander},
  journal={Advances in neural information processing systems},
  volume={34},
  pages={8780--8794},
  year={2021}
}

@article{ho2022classifier,
  title={Classifier-free diffusion guidance},
  author={Ho, Jonathan and Salimans, Tim},
  journal={arXiv preprint arXiv:2207.12598},
  year={2022}
}

@article{zhao2024probabilistic,
  title={Probabilistic inference in language models via twisted sequential monte carlo},
  author={Zhao, Stephen and Brekelmans, Rob and Makhzani, Alireza and Grosse, Roger},
  journal={arXiv preprint arXiv:2404.17546},
  year={2024}
}

@article{davis2024fisher,
  title={Fisher flow matching for generative modeling over discrete data},
  author={Davis, Oscar and Kessler, Samuel and Petrache, Mircea and Ceylan, Ismail Ilkan and Bronstein, Michael and Bose, Avishek Joey},
  journal={arXiv preprint arXiv:2405.14664},
  year={2024}
}

@article{miller2024flowmm,
  title={FlowMM: Generating Materials with Riemannian Flow Matching},
  author={Miller, Benjamin Kurt and Chen, Ricky TQ and Sriram, Anuroop and Wood, Brandon M},
  journal={arXiv preprint arXiv:2406.04713},
  year={2024}
}

@article{akhound2024iterated,
  title={Iterated denoising energy matching for sampling from Boltzmann densities},
  author={Akhound-Sadegh, Tara and Rector-Brooks, Jarrid and Bose, Avishek Joey and Mittal, Sarthak and Lemos, Pablo and Liu, Cheng-Hao and Sendera, Marcin and Ravanbakhsh, Siamak and Gidel, Gauthier and Bengio, Yoshua and others},
  journal={arXiv preprint arXiv:2402.06121},
  year={2024}
}

@article{wang2022data,
  title={From data to noise to data for mixing physics across temperatures with generative artificial intelligence},
  author={Wang, Yihang and Herron, Lukas and Tiwary, Pratyush},
  journal={Proceedings of the National Academy of Sciences},
  volume={119},
  number={32},
  pages={e2203656119},
  year={2022},
  publisher={National Acad Sciences}
}

@article{kadanoff1966scaling,
  title={Scaling laws for Ising models near $T_c$},
  author={Kadanoff, Leo P},
  journal={Physics Physique Fizika},
  volume={2},
  number={6},
  pages={263},
  year={1966},
  publisher={APS}
}

@article{wilson1971renormalization1,
  title={Renormalization group and critical phenomena. I. Renormalization group and the Kadanoff scaling picture},
  author={Wilson, Kenneth G},
  journal={Physical review B},
  volume={4},
  number={9},
  pages={3174},
  year={1971},
  publisher={APS}
}

@article{wilson1971renormalization2,
  title={Renormalization group and critical phenomena. II. Phase-space cell analysis of critical behavior},
  author={Wilson, Kenneth G},
  journal={Physical Review B},
  volume={4},
  number={9},
  pages={3184},
  year={1971},
  publisher={APS}
}

@article{wilson1983renormalization,
  title={The renormalization group and critical phenomena},
  author={Wilson, Kenneth G},
  journal={Reviews of Modern Physics},
  volume={55},
  number={3},
  pages={583},
  year={1983},
  publisher={APS}
}

@article{dibak2022temperature,
  title={Temperature steerable flows and Boltzmann generators},
  author={Dibak, Manuel and Klein, Leon and Kr{\"a}mer, Andreas and No{\'e}, Frank},
  journal={Physical Review Research},
  volume={4},
  number={4},
  pages={L042005},
  year={2022},
  publisher={APS}
}

@article{davis1959leonhard,
  title={Leonhard euler's integral: A historical profile of the gamma function: In memoriam: Milton abramowitz},
  author={Davis, Philip J},
  journal={The American Mathematical Monthly},
  volume={66},
  number={10},
  pages={849--869},
  year={1959},
  publisher={Taylor \& Francis}
}

@article{nicoli2020asymptotically,
  title={Asymptotically unbiased estimation of physical observables with neural samplers},
  author={Nicoli, Kim A and Nakajima, Shinichi and Strodthoff, Nils and Samek, Wojciech and M{\"u}ller, Klaus-Robert and Kessel, Pan},
  journal={Physical Review E},
  volume={101},
  number={2},
  pages={023304},
  year={2020},
  publisher={APS}
}

@article{nicoli2021estimation,
  title={Estimation of thermodynamic observables in lattice field theories with deep generative models},
  author={Nicoli, Kim A and Anders, Christopher J and Funcke, Lena and Hartung, Tobias and Jansen, Karl and Kessel, Pan and Nakajima, Shinichi and Stornati, Paolo},
  journal={Physical review letters},
  volume={126},
  number={3},
  pages={032001},
  year={2021},
  publisher={APS}
}

@article{bulgarelli2024flow,
  title={Flow-based Sampling for Entanglement Entropy and the Machine Learning of Defects},
  author={Bulgarelli, Andrea and Cellini, Elia and Jansen, Karl and K{\"u}hn, Stefan and Nada, Alessandro and Nakajima, Shinichi and Nicoli, Kim A and Panero, Marco},
  journal={arXiv preprint arXiv:2410.14466},
  year={2024}
}

@article{singha2025multilevel,
  title={Multilevel Generative Samplers for Investigating Critical Phenomena},
  author={Singha, Ankur and Cellini, Elia and Nicoli, Kim A and Jansen, Karl and K{\"u}hn, Stefan and Nakajima, Shinichi},
  journal={arXiv preprint arXiv:2503.08918},
  year={2025}
}

@article{herron2024inferring,
  title={Inferring phase transitions and critical exponents from limited observations with thermodynamic maps},
  author={Herron, Lukas and Mondal, Kinjal and Schneekloth Jr, John S and Tiwary, Pratyush},
  journal={Proceedings of the National Academy of Sciences},
  volume={121},
  number={52},
  pages={e2321971121},
  year={2024},
  publisher={National Academy of Sciences}
}

@article{kingma2014adam,
  title={Adam: A method for stochastic optimization},
  author={Kingma, Diederik P and Ba, Jimmy},
  journal={arXiv preprint arXiv:1412.6980},
  year={2014}
}
\addcontentsline{toc}{section}{References}

\clearpage

\appendix
\renewcommand{\thesection}{\Roman{section}}
\renewcommand{\thefigure}{S\arabic{figure}}
\setcounter{figure}{0}

\begin{center}
{\LARGE \bfseries Supporting Information \par}
\end{center}
\section*{}
\addcontentsline{toc}{section}{Supporting Information}

\section{Notations}

\subsection{Notations for the Spin States}

The discrete spin state is denoted by $s \in \left\{ -1, 1 \right\}$, and $s_i$ denotes the discrete state of spin $i$.

Meanwhile, we also use a vector of continuous variables $\boldsymbol{x} =(x_{(0)},x_{(1)}) \in \left[ 0, 1 \right]^2$ with $x_{(0)}+x_{(1)}=1$ to denote the spin state in terms of the probabilities of the two-class categorical distribution. It is also referred to as a two-class simplex. And we use $\boldsymbol{x}_i$ to denote the continuous state of spin $i$. The flow matching algorithm is built based on the continuous spin states.

We use $\hat{x}$ to denote the state of an Ising lattice with $N$ spins.

\subsection{Notations Used in the Flow Matching Algorithms}

Throughout, we denote the probability density function of the random prior as $q$ and the probability density function of the target data as $p_\mathrm{data}$.

We denote the conditional probability path by $p_t \left( \boldsymbol{x} \mid \boldsymbol{x} \left( 1 \right) \right)$ and the marginal probability path by $p_t \left( \boldsymbol{x} \right)$. The corresponding conditional and marginal velocity fields are written as $\boldsymbol{u}_t \left( \boldsymbol{x} \mid \boldsymbol{x} \left( 1 \right) \right)$ and $\boldsymbol{v}_t \left( \boldsymbol{x} \right)$, respectively. Here, $\boldsymbol{x}$ represents the spin state at time $t\in \left[ 0, 1 \right]$, and $\boldsymbol{x} \left( 1 \right)$ denotes the spin state at time $t=1$. The variable $t$ refers to the fictitious flow-matching time and carries no physical interpretation.

The probability of a lattice configuration is denoted by $P \left( \hat{x} \right)$. 

The dependence on $t$ may be expressed either as $p_t$ or as $p \left( \boldsymbol{x} ; t \right)$ when potential conflicts with other subscripts arise.

Throughout the paper, we denote scores by $\boldsymbol{\epsilon}$ and flows by $\boldsymbol{u}$.

\subsection{Notations Used for the Model Architecture}

We denote the energy of a lattice configuration as $E$ and the magnetization as $m$.

We implement message passing using convolutional layers. At each layer $l$, we define $A_i^{(l)}$ as the node representation, $\mathcal{T}^{(l)}$ as the representation of time, and $\mathcal{E}^{(l)}$ and $\mathcal{M}^{(l)}$ as the representations of conditions $E$ and $m$, respectively.

The message passing update at layer $l$ is given by $M_i^{(l)}=\sum_{j \in \mathcal{N}(i) \cup \{i\}} H^{(l)}(\vec{r}_{ji})(A^{(l)}_{j}+\mathcal{T}^{(l)}+\mathcal{E}^{(l)}+\mathcal{M}^{(l)})$ where $H^{(l)}(\vec{r}_{ji})$ consists of nine matrices of dimension $N_\mathrm{embedding}\times N_\mathrm{embedding}$, corresponding to the eight neighbors considered for each spin, along with the self-interaction. 

$\boldsymbol{g}_i$ denotes the model output for spin $i$, and $\hat{g}$ denotes the model output for an Ising lattice with $N$ sites.

We denote loss functions with $\mathcal{L}$. Three different loss functions are used, namely $\mathcal{L}_\mathrm{CE}$, $\mathcal{L}_\mathrm{E}$, and $\mathcal{L}_\mathrm{RC}$.

\clearpage

\section{Convergence of the Dirichlet Probability Path over Integration time}
\label{app:conv-dir}
\begin{figure}[h]
    \centering
    \includegraphics[width=1\textwidth]{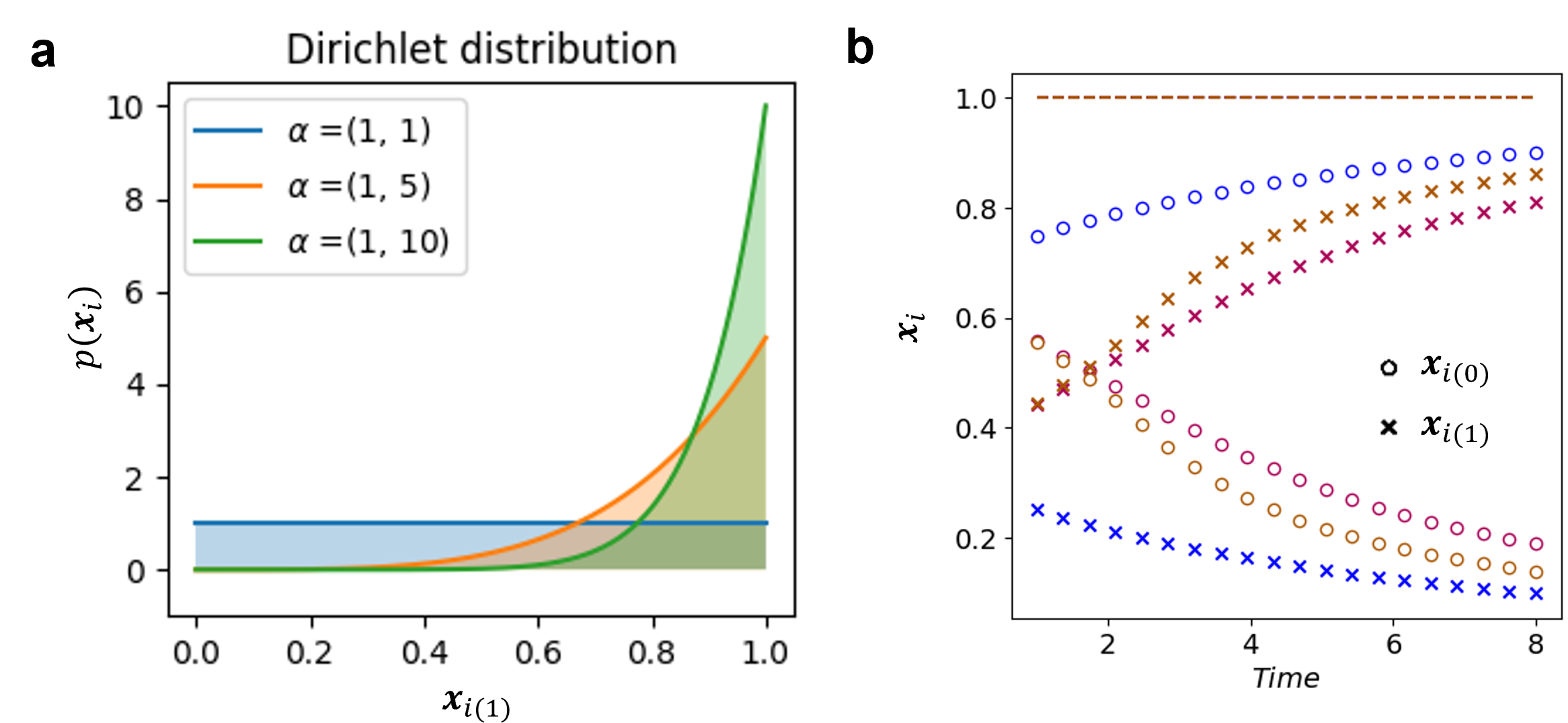}
    \caption{(a) Dirichlet distribution with various $\boldsymbol{\alpha}$ parameters. (b) Convergence of the Dirichlet probability path over integration time. Three examples are given, each in a distinct color.}
    \label{fig:conv-dirichlet}
\end{figure}

\clearpage

\section{Algorithms of Training and Sampling of the Flow Matching Models}
Algorithms \ref{algo1} and \ref{algo2} describe in detail how to train the flow matching models and how to sample using the trained flow matching models.

\begin{algorithm}[H]
\caption{Training the Flow Model Using the Cross-Entropy Loss}
\label{algo1}

\KwIn{Training data $\hat{x}(1) = \left\{ \boldsymbol{x}_i(1); i = 1, 2, \dots, N \right\}$}
\KwOut{Trained model $f_\theta \left( \boldsymbol{x}_i, \left\{ \boldsymbol{x}_j \right\}_{j\in \mathcal{N}(i)}; t \right)$, where $\theta$ denote learnable model parameters}

Initialize $f_\theta$ with random parameters $\theta$\;

\Repeat{
    $-\lambda_{\mathrm{CE}}\mathbb{E}_t \sum_i \left[ \boldsymbol{x}_{i}(1) \cdot \ln \boldsymbol{g}_i (t) \right] \mathrm{\ has \ converged}$
}{
    Sample $\boldsymbol{x}_{i}(0) \sim \mathrm{Dir} \left( \boldsymbol{\alpha} = (1,1) \right)$\;
    Sample $t \sim \mathrm{Exp}(0.5)$\;
    Sample $\boldsymbol{x}_i \sim \mathrm{Dir} \left( \boldsymbol{\alpha} = (1,1) + \boldsymbol{x}_{i}(1) \cdot t \alpha_\mathrm{max} \right)$\;
    Compute classifier: $\boldsymbol{g}_{i} \gets f_\theta \left( \boldsymbol{x}_i, \left\{ \boldsymbol{x}_j \right\}_{j\in \mathcal{N}(i)}; t \right)$
    
    Take gradient descent step on $-\lambda_{\mathrm{CE}}\mathbb{E}_t \sum_i \left[ \boldsymbol{x}_{i}(1)\cdot \ln \boldsymbol{g}_i(t) \right]$\;
}

\end{algorithm}
Here, we use $\boldsymbol{x}_{i}(1)$ to denote the target state of spin $i$, and $\boldsymbol{x}_{i}(0)$ to denote the initial random state of spin $i$. We illustrate the method using the cross-entropy loss; training with other loss functions follows analogously.

\begin{algorithm}[H]
\caption{Flow Matching Generation Process}
\label{algo2}

\KwIn{Trained flow model $f_\theta \left( \boldsymbol{x}_i, \left\{ \boldsymbol{x}_j \right\}_{j\in \mathcal{N}(i)}; t \right)$}
\KwIn{Initial sample $\boldsymbol{x}_{i}(0) \sim \mathrm{Dir} \left( \boldsymbol{\alpha}=(1,1) \right)$}
\KwIn{Time discretization $\left\{ t_n \right\}_{n=0}^\xi$}
\KwOut{Generated sample $\boldsymbol{x} \left( t_\xi \right)$}

Initialize $\boldsymbol{x}_{i}(0)$\;

\For{$n \gets 1$ \KwTo $\xi$}{
    Compute classifier: $\boldsymbol{g}_i \gets f_\theta \left( \boldsymbol{x}_i, \left\{ \boldsymbol{x}_j \right\}_{j\in \mathcal{N}(i)}; t_n \right)$\;
    Compute velocity: $\boldsymbol{v}_i \left( t_n \right) \gets g_{i(0)} \boldsymbol{u}_{t_n} \left( \boldsymbol{x}_i \mid \boldsymbol{x}_i(1) = (1,0) \right) + g_{i(1)} \boldsymbol{u}_{t_n} \left( \boldsymbol{x}_i \mid \boldsymbol{x}_i(1) = (0,1) \right)$\;
    Update state: $\boldsymbol{x}_i \left( t_{n+1} \right) \gets \boldsymbol{x}_i \left( t_n \right) + \boldsymbol{v}_i \left( t_n \right) \left( t_n - t_{n-1} \right)$\;
}
\Return $\boldsymbol{x}_i \left( t_\xi \right)$

\end{algorithm}

\clearpage

\section{Algorithm of Guided Generation for Multiple Temperatures}

Algorithm \ref{algo3} describes the guided sampling for multiple temperatures.

\begin{algorithm}[H]
\caption{Multitemperature Generation Process}
\label{algo3}

\KwIn{Trained flow model $f_\theta \left( \boldsymbol{x}_i, \left\{\boldsymbol{x}_j \right\}_{j\in \mathcal{N}(i)}; t \right)$, conditional flow model $f^{\prime}_\theta \left( \boldsymbol{x}_i, \left\{\boldsymbol{x}_j \right\}_{j\in \mathcal{N}(i)};t \mid m, E \right)$}
\KwIn{Initial state $\boldsymbol{x}_{i}(0) \sim \mathrm{Dir} \left( \boldsymbol{\alpha}=(1,1) \right)$}
\KwIn{Time discretization $\left\{ t_n \right\}_{n=0}^\xi$, guidance strength $\gamma \left( T \right)$}
\KwIn{$m$, $E$ of the zero temperature magnetic state of $6\times6$ Ising model}
\KwOut{Final generated samples $\boldsymbol{x}_i \left( t_\xi \right)$}

Initialize $\boldsymbol{x}_{i}(0)$\;

\For{$n \gets 1$ \KwTo $\xi$}{
    Compute classifier: $\boldsymbol{g}_i \gets f_\theta \left( \boldsymbol{x}_i, \left\{ \boldsymbol{x}_j \right\}_{j\in \mathcal{N}(i)}; t_n \right)$\;
    Compute conditional classifier: $\boldsymbol{g}^{\prime}_i \gets f^{\prime}_\theta \left( \boldsymbol{x}_i, \left\{ \boldsymbol{x}_j \right\}_{j\in \mathcal{N}(i)}; t_n \mid m, E \right)$\;
    Compute guided classifier: $\boldsymbol{g}_{\mathrm{CFG},i} \gets \left( \boldsymbol{g}_i^{\prime} \right)^\gamma * \boldsymbol{g}_i^{1-\gamma}$\;
    Compute velocity: $\boldsymbol{v}_i \left( t_n \right) \gets g_{\mathrm{CFG},i(0)}\boldsymbol{u}_{t_n} \left( \boldsymbol{x}_i \mid \boldsymbol{x}_i(1) = (1,0) \right) + g_{\mathrm{CFG},i(1)}\boldsymbol{u}_{t_n}\left( \boldsymbol{x}_i \mid \boldsymbol{x}_i(1) = (0,1) \right)$\;
    Update state: $\boldsymbol{x}_i \left( t_{n+1} \right) \gets \boldsymbol{x}_i(t_n) + \boldsymbol{v}_i \left( t_n \right) \left( t_n - t_{n-1} \right)$\;
}
\Return $\boldsymbol{x}_i \left( t_\xi \right)$

\end{algorithm}

\clearpage

\section{Guidance Strength \texorpdfstring{$\gamma(T)$}{}}
\begin{figure}[h]
    \centering
    \includegraphics[width=0.5\linewidth]{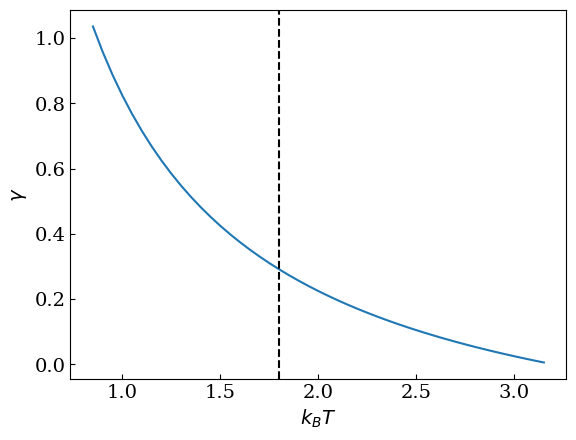}
    \caption{The guidance strength $\gamma$ as a function of temperature.}
    \label{fig:gamma-T}
\end{figure}

The guidance strength $\gamma(T)$ at different temperatures is given in Figure~\ref{fig:gamma-T}.

\clearpage

\section{Loss Prefactors and Training Cost}
\label{app:training-cost}
\begin{figure}[h]
    \centering
    \includegraphics[width=1\linewidth]{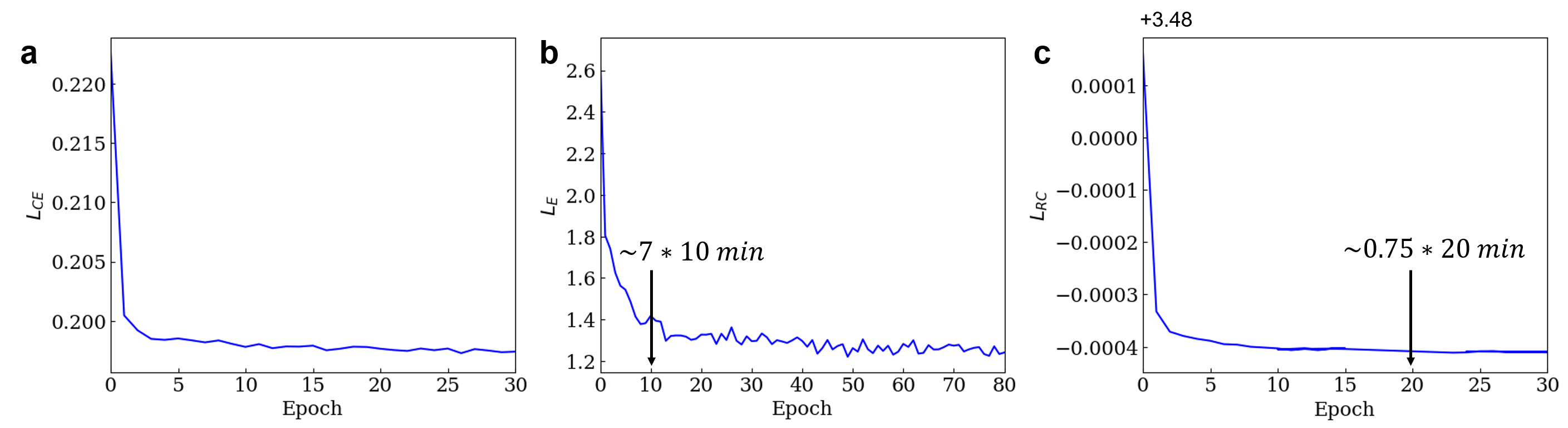}
    \caption{Convergence line of loss functions (a) $\mathcal{L}_{\mathrm{CE}}$, (b) $\mathcal{L}_{\mathrm{E}}$, and (c) $\mathcal{L}_{\mathrm{RC}}$. And the GPU time on H100 for combined training of $\mathcal{L}_{\mathrm{E}}+\mathcal{L}_{\mathrm{CE}}$ and $\mathcal{L}_{\mathrm{RC}}+\mathcal{L}_{\mathrm{CE}}$ respectively. Since $\mathcal{L}_{\mathrm{E}}$ is relatively expensive to calculate, we only use it for 10 epochs to pretrain the model, and use $\mathcal{L}_{\mathrm{RC}}+\mathcal{L}_{\mathrm{CE}}$ for further convergence.}
    \label{fig:losses}
\end{figure}

We begin by pretraining the flow model using a combination of $\mathcal{L}_{\mathrm{CE}}$ and $\mathcal{L}_{\mathrm{E}}$, with prefactors $\lambda_{\mathrm{CE}}=1$, $\lambda_{\mathrm{E}}=1$, $\lambda_{\mathrm{MAE}}=1$ and $\sigma=500$. Calculating the energy loss is relatively expensive, so we only use it for 10 epochs. 
Typically, both loss functions decrease significantly within the first 5 epochs.
Following this initial phase, the flow model was further converged using $\mathcal{L}_{\mathrm{CE}}$ and $\mathcal{L}_{\mathrm{RC}}$, where the magnetization $m=\sum_{i=1}^{N} s_i$ is used as reaction coordinate and the prefactors are $\lambda_{\mathrm{RC}}=10$, $\lambda_{\mathrm{CE}}=1$. $\mathcal{L}_{\mathrm{RC}}$ and $\mathcal{L}_{\mathrm{CE}}$ converge after around $20$ epochs. 
In Figure \ref{fig:losses}, we provide the convergence lines of different loss functions and the corresponding GPU time.

Note that the high cost of the energy loss here can be reduced through parallelization. But since we only need it for 10 epochs, we did not implement parallelization in our code.

\clearpage

\section{Heat capacity of the Ising model}

\begin{figure}[h]
    \centering
    \includegraphics[width=1\linewidth]{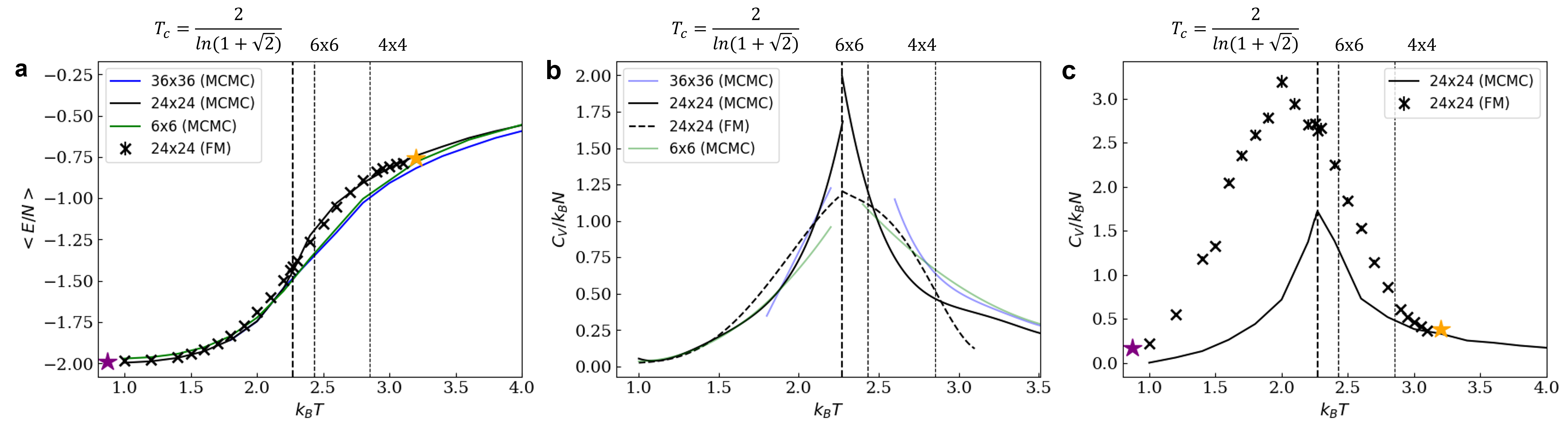}
    \caption{(a) The expected potential energy per spin $\langle E/N \rangle$ at various temperatures predicted by guided flow matching for a $24 \times 24$ lattice (crosses), compared against MCMC data (lines). (b) Heat capacity of the $24 \times 24$ lattice computed from the temperature derivative $\delta\langle E \rangle/\delta T$. (c) Heat capacity of the $24 \times 24$ lattice computed using the fluctuation formula. The vertical lines indicate the critical temperatures of phase transition for different lattice sizes obtained by the renormalization group theory~\cite{wilson1971renormalization1,wilson1971renormalization2,wilson1983renormalization}.}
    \label{fig:heat-capacity}
\end{figure}

To further assess the effectiveness of the flow matching model, we computed the heat capacity of a $24\times 24$ lattice using two complementary approaches.
(1) The first method computes $C_V$ as the derivative of the mean energy with respect to temperature, 
\begin{equation}
C_V=\frac{\delta\langle E\rangle}{\delta T}
,
\end{equation}
with results shown in Figure~\ref{fig:heat-capacity}(b). 
(2) Alternatively, exploiting the fluctuation-dissipation relation in the canonical ensemble, the heat capacity can be expressed as
\begin{equation}
    C_V=\frac{1}{k_BT^2} \left( \left\langle E^2 \right\rangle - \left\langle E \right\rangle^2 \right)
    ,
\end{equation}
as plotted in Figure~\ref{fig:heat-capacity}(c). 

\begin{figure}
    \centering
    \includegraphics[width=0.5\linewidth]{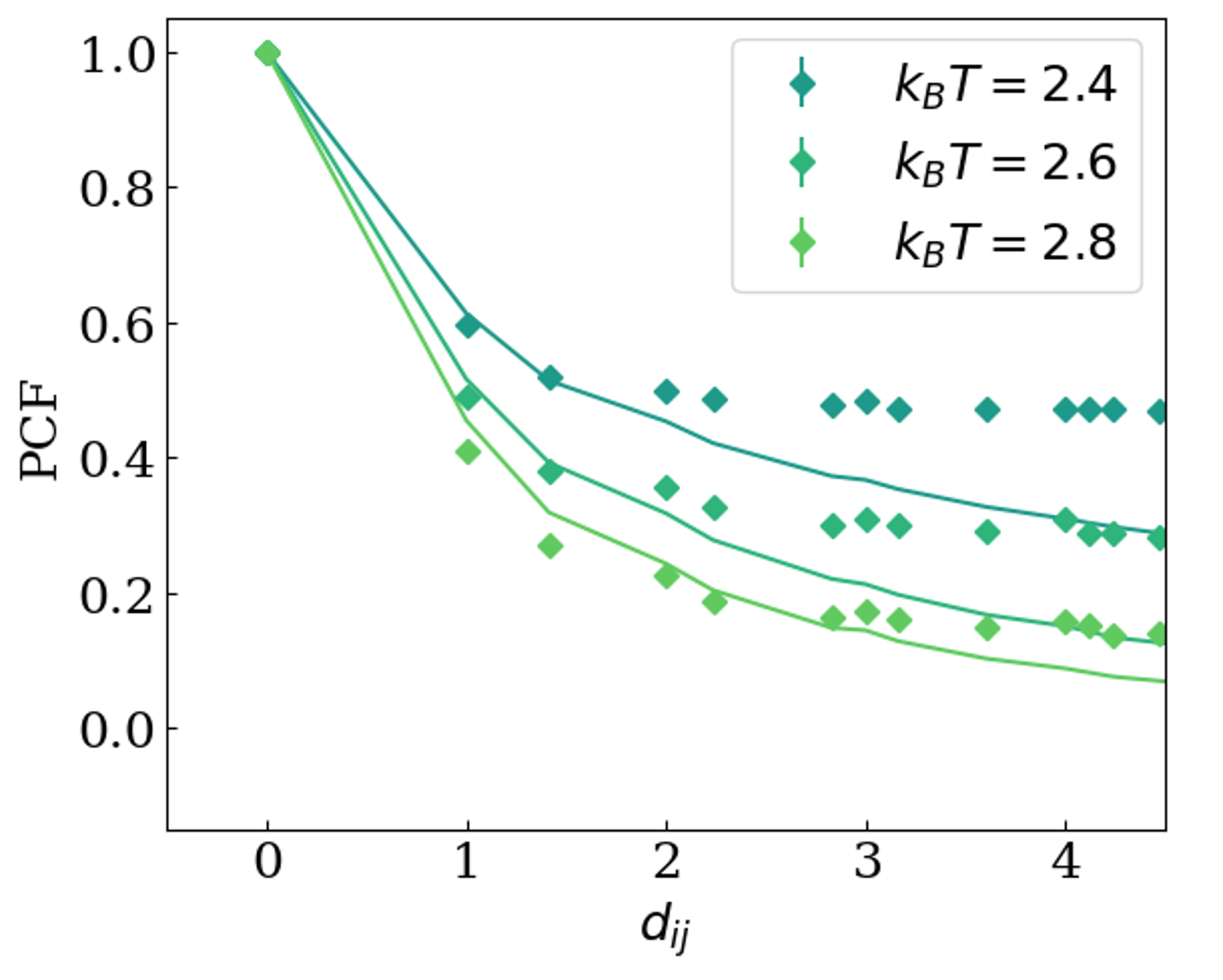}
    \caption{Pair correlation function of a $24\times 24$ lattice at temperatures near $T_c$.}
    \label{fig:pcf-Tc}
\end{figure}

The derivative-based estimate yields a smooth curve, with underestimated heat capacities at $T\sim T_c$, as expected. Figure~\ref{fig:pcf-Tc} gives the pair correlation function at temperatures near $T_c$, where the gradient of the free energy diverges and is not captured by the flow matching model. 

However, the fluctuation-based estimate in Figure~\ref{fig:heat-capacity}(c) exhibits a systematic overestimation of $C_V$, which is consistent with the presence of large energy outliers produced by the model, as shown in Figure~\ref{fig:2s2-guided} of the main text.

\clearpage

\section{Finite Size Effect of 2D Lattice Ising Model}
\label{app:finite-size-effect}
\begin{figure}[h]
    \centering
    \includegraphics[width=1.0\linewidth]{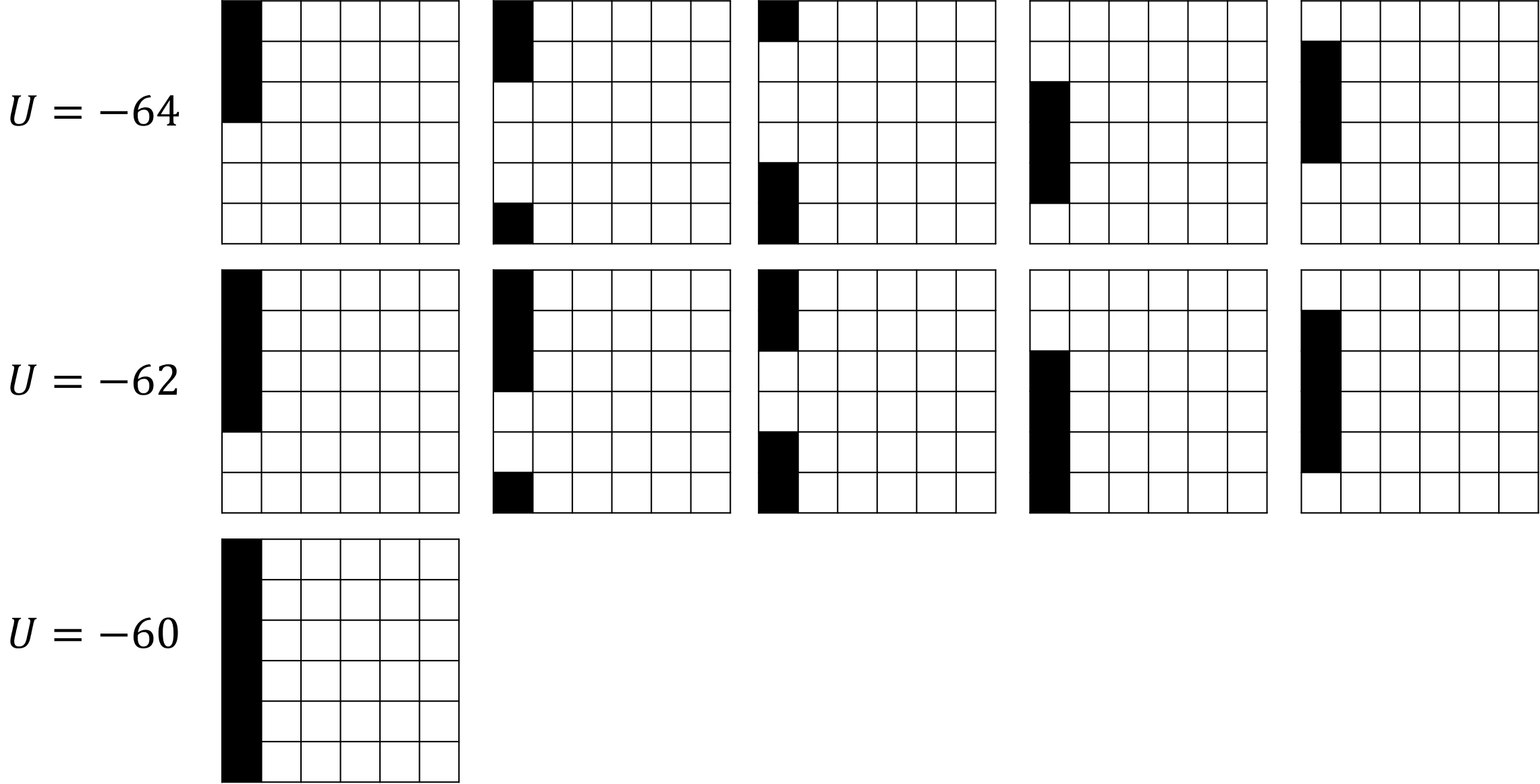}
    \caption{Enumeration of all possible configurations of $6\times 6$ lattice Ising model with a single 1D domain located at the leftmost edge of the lattice.}
    \label{fig:1d-domain}
\end{figure}

Here, we illustrate how the finite-size effect of a $6 \times 6$ lattice Ising model can yield a higher free energy for configurations with $E/N \approx -1.667$. The minimum potential energy of a $6\times 6$ Ising model is $-72$ of magnetic states, with either all positive spins or all negative spins.  While at $E=-60$, the free energy curve shows a bump. This is because the domain wall meets the lattice boundary when $E=-60$. To analyze this, we enumerate all possible configurations under a strict condition: there must be a single 1D domain located at the leftmost edge of the lattice. Figure~\ref{fig:1d-domain} presents the results, grouping configurations by their potential energies $E$. We find that for $E < -60$, there are five distinct configurations, whereas for $E = -60$, where the domain wall coincides with the lattice boundary, only one configuration is possible. 
Consequently, the fewer states at $E = -60$ correspond to a higher free energy.

\end{document}